**Tunable Ultrastrong Magnon-Magnon Coupling and Squeezed States in a Triaxial Van der Waals Antiferromagnet**

Xiao Xiao[1], Zishuang Li[1], Zui Tao[1], Lina Chen[2,3], Ping Liu[2], Shuai Zhang[1], Haozhe Wang[1], Yong Pu[2] and Ronghua Liu[1,4]*

[1] National Key Laboratory of Spintronics and School of Physics, Nanjing University, Nanjing 210093, China

[2] School of Science, Nanjing University of Posts and Telecommunications, Nanjing 210023, China

[3] Laboratory of Solid State Microstructures, Nanjing University, Nanjing 210093, China

[4] State Key Laboratory of Spintronics, Nanjing University, Suzhou 215163, China

*Corresponding Email: rhliu@nju.edu.cn

**Abstract:**

Achieving and quantifying ultrastrong magnon-magnon coupling (USC) remains a key challenge for quantum magnonics. Here, we demonstrate widely tunable USC between chiral (right-/left-handed) and acoustic/optical magnon modes in the van der Waals antiferromagnet $CrPS_4$. Using a full quantum model, we decompose the interaction into three microscopic interaction channels: the co-rotating term $g_1$ (mode hybridization), the counter-rotating term $g_2$ (two-mode squeezing), and, uniquely in the chiral regime, the self-squared term $g_3$ that mediates single-mode squeezing. The normalized coupling ratio reaches $g/f_g$~0.41, entering the USC regime. Most significantly, we establish a complete protocol linking these USC parameters to a measurable quantum phenomenon: a squeezed magnon vacuum state. Our simulations not only visualize the chirality and polarization evolution of the hybrid modes but also demonstrate how to directly extract quadrature fluctuations and a substantial squeezing factor (up to ~9.7 dB) from thermal noise-driven magnetization dynamics. This work provides a material-specific quantum framework for USC and a practical pathway to detect and exploit non-classical magnon states in low-damping two-dimensional magnets, opening a route toward quantum-enhanced spintronics and hybrid quantum information platforms.

Quantum magnonics, which explores coherent magnon-matter interactions for low-power quantum information processing and storage, has attracted significant interest in recent years [1–4]. Within hybrid magnonic systems, magnon-magnon coupling represents a fundamental interaction in magnetic materials and plays a crucial role in spintronics and quantum information science [4–7]. Tunable magnon-magnon coupling has been experimentally demonstrated in a variety of magnetic systems, including magnetic multilayers [8,9], synthetic antiferromagnets (SAFs) [4,6,10–17], and two-dimensional (2D) van der Waals (vdW) antiferromagnets (AFMs), such as $CrCl_3$ [7,18], CrSBr [19,20], $MnPS_3$ [21], $CrPS_4$ [22–24].

A key challenge arises when the coupling strength $g$ becomes comparable to the magnon resonance frequencies $f_g$, entering the strong and ultrastrong coupling (USC) regimes. In this regime, conventional methods for quantifying coupling, such as the minimal frequency splitting (direct gap) or field-dependent frequency difference (indirect gap) [7,13,24], break down. These approaches typically rely on the rotating-wave approximation (RWA) [25], which neglects counter-rotating terms responsible for non-energy-conserving processes like simultaneous magnon pair creation/annihilation [4,28–31]. The RWA is valid only when $g \ll f_g$; as $g/f_g$ approaches or exceeds 0.1, counter-rotating interactions induce significant spectral shifts (e.g., the vacuum Bloch-Siegert shift) and reshape the quantum ground state. These interactions also generate squeezed magnon states, where quantum fluctuations in one quadrature are suppressed below the standard quantum limit (SQL) [32,33]. Such entangled, non-classical states are a direct resource for quantum-enhanced sensing and information processing. However, the experimental identification of these squeezed states and their quantitative linkage to USC parameters through directly measurable spin dynamics in a real magnetic material, particularly in 2D vdW magnets, remains an outstanding challenge.

The A-type vdW AFM $CrPS_4$ presents a uniquely promising platform to address this challenge. Its low crystal symmetry gives rise to a triaxial magnetic anisotropy, combining an out-of-plane monoclinic easy *c*-axis with in-plane orthorhombic anisotropies ($\beta = 91.88°$ between *a* and *c*-axis, with *b* perpendicular to *a* and *c* axes) [34–38], and supports a rich spectrum of magnon modes: acoustic (AM) and optical (OM) modes, as well as right- and left-handed chiral modes (RM and LM). Together with weak interlayer exchange, this anisotropy yields accessible GHz-range antiferromagnetic resonances (AFMR) and tunable hybrid magnon modes under temperature and magnetic field [22]. Despite this potential, prior studies of magnon coupling in $CrPS_4$ have been confined to classical models within the RWA [22–24], leaving its quantum USC physics and the associated non-classical magnon states entirely unexplored.

In this Letter, we go beyond the RWA to unveil the quantum physics of magnon-magnon USC in $CrPS_4$. By experimentally controlling two distinct USC types (AM-OM and RM-LM) via temperature and field orientation, we develop a quantum model beyond the RWA that decomposes the interaction into three microscopic channels: the co-rotating term $g_1$ (hybridization), the counter-rotating term $g_2$ (two-mode squeezing), and, crucially in the RM-LM regime, the self-squared term $g_3$, which corresponds to a previously unreported single-mode squeezing mechanism. We quantitatively extract all three coupling strengths from spectra, with a normalized ratio $g/f_g$ reaching 0.41. Most importantly, we establish a general, experimentally actionable protocol to connect USC to quantum squeezing. By combining our quantum model with micromagnetic simulations, we visualize the chirality and polarization evolution of the hybrid modes and extract quadrature fluctuations with a squeezing factor up to ~9.7 dB from simulated magnetization dynamics.

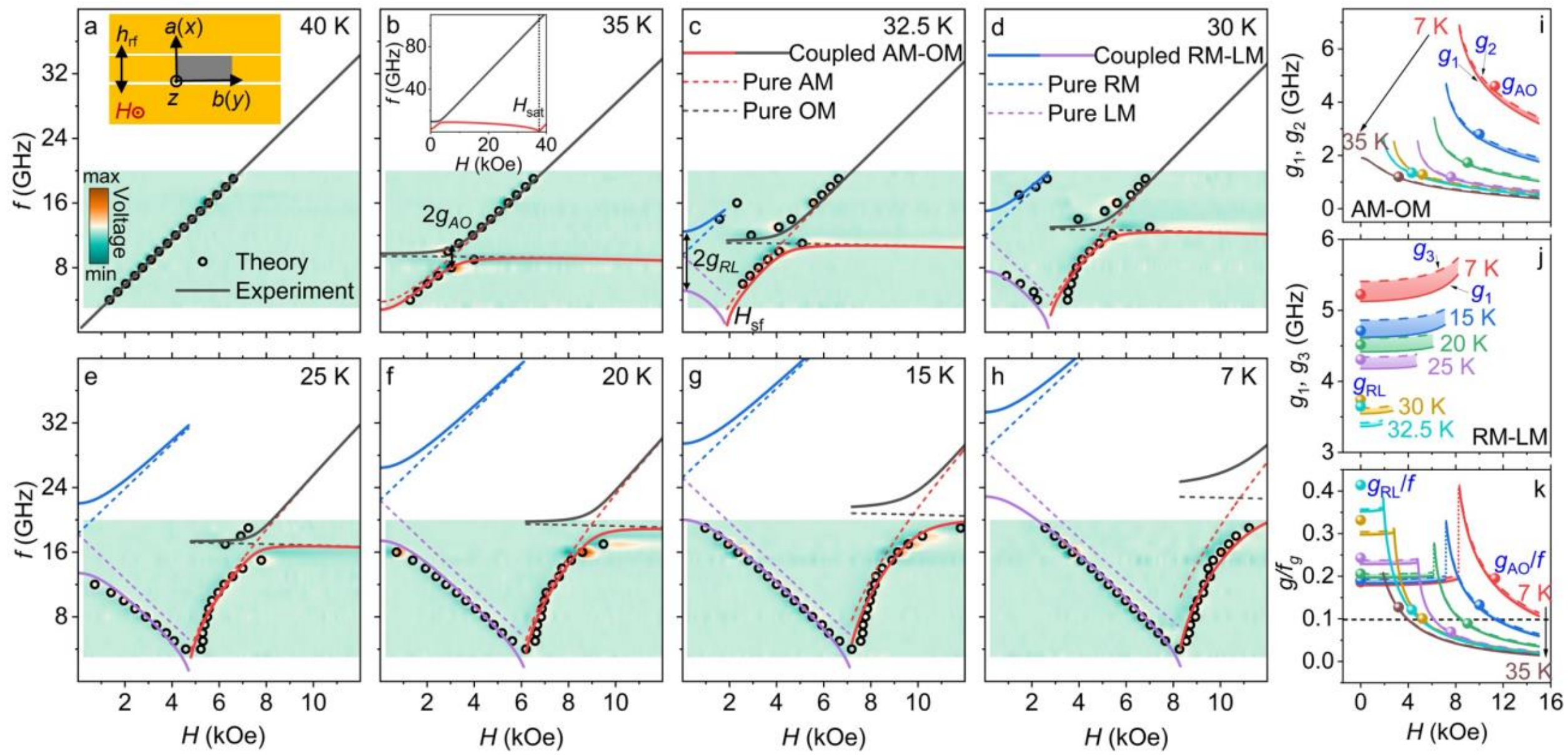

**Figure 1.** Experimental spectra and theoretically calculated frequency dispersions and coupling strength of the AFMR modes at different temperatures under $H \parallel z$. (a–h) Color maps of the experimental differential AFMR spectra (d$P$/d$H$) acquired from 3-20 GHz at the labeled temperatures. Black open circles denote the resonance fields extracted from the d$P$/d$H$ spectra. The red/black (blue/purple) dashed line denotes the theoretical dispersion of the pure AM/OM (RM/LM) without considering the mode coupling, while the red-black (blue-purple) solid lines show the theoretically fitted dispersion of the coupled AM-OM (RM-LM) hybrid mode. The spin-flop field $H_{sf}$ is the boundary between the RM-LM and AM-OM regimes. Inset of (a): the experimental setup; the vertical dotted line in (b) inset represents the saturation field $H_{sat}$. (i–k) Field dependence of coupling strengths extracted from the quantum model at the labeled temperatures in (a–h). Solid lines: $g_1$, dashed lines: $g_2$ (i) and $g_3$ (j). Spherical symbols indicate half the minimum frequency gap $g_{AO}$ and $g_{RL}$ directly extracted from the AFMR frequency dispersion spectra. (k) Summary of the normalized coupling ratio $g/f_g$, where $f_g = (f_1+f_2)/2$ is the average frequency of two modes at a given magnetic field.

**AFMR dispersion spectra:** The differential AFMR measurements were performed on a 4 × 0.7 × 0.1 mm $CrPS_4$ single crystal mounted on a coplanar waveguide (CPW) [40]. The crystal was oriented with its crystallographic *b*-axis (the longer dimension) aligned along the signal line to optimize signal detection [inset of Fig. 1(a)] [39]. At $T$ = 40 K, above the Néel temperature ($T_N$ = 36.5 K, as determined from static magnetization measurements; see Section I of the Supplemental Material (SM) [40]), the system exhibits a paramagnetic (PM) state with short-range magnetic order, displaying a single resonance peak [Fig. 1(a)]. The resonance frequency follows the Kittel formula for FMR, $f = \gamma(H - M_{eff})$, with a gyromagnetic ratio $\gamma$ = 2.84 GHz/kOe and an effective magnetization $M_{eff}$ = -0.05 kOe.

Upon cooling to 35 K, still above the spin-reorientation transition temperature $T^*$ = 32.5 K, the system enters the $AFM_b$ phase, where spins remain antiparallel along the $b$-axis [40]. Here, two distinct resonance modes emerge. The low-frequency branch initially increases with applied field $H$ but softens to zero at higher fields. The high-frequency branch rises continuously before saturating at $H_{sat}$ = 37.5 kOe [inset of Fig. 1(b)]. The frequency-dispersion spectra [Fig. 1(b)] reveal an anticrossing behavior near 3 kOe, indicative of an AM-OM hybrid mode.

At $T = T^*$ = 32.5 K, the easy axis reorients from the $b$-axis to the $c$-axis ($AFM_c$ state), introducing an additional RM-LM hybrid mode at low fields, coexisting with the AM-OM mode observed at 35 K. The RM-LM mode exhibits a pronounced zero-field frequency gap, reflecting the in-plane orthogonal anisotropy between the $a$- and $b$-axes. A spin-flop transition occurs at $H_{sf}$ = 1.9 kOe, where the $Cr^{2+}$ spins reorient from antiparallel alignment along the $c$-axis back to the $b$-axis, accompanied by a dynamic transition from the RM-LM to the AM-OM hybrid mode. Below $T^*$, the AFMR spectra maintain similar characteristics across temperatures, with only an overall frequency shift observed.

**Classical LL model:** To quantitatively describe the observed magnon dispersions in $CrPS_4$, we first employ a classical Landau-Lifshitz (LL) model (see Section II of SM [40] for details). The model yields excellent agreement with the measured dispersion relations. Temperature-dependent fitting parameters, including the effective anisotropy fields ($H_{A1}$, $H_{A2}$, $H_{Am}$) and the exchange field ($H_E$), are summarized in Fig. S7(b) [40]. For $T^* < T < T_N$, the inequality $H_{A2} < H_{Am} < H_{A1}$ is consistent with the $AFM_b$ ground state. Below $T^*$ = 32.5 K, the relation becomes $H_{Am} < H_{A2} < H_{A1}$, reflecting the expected $AFM_c$ ground state.

To isolate the effect of mode coupling, we first examine the dispersion of the uncoupled spin-wave modes. In the AM-OM coupling regime, the in-phase precession of the two sublattice magnetizations $\boldsymbol{m}_1$ and $\boldsymbol{m}_2$ defines the AM, labeled $\delta\boldsymbol{m}_+$, while the out-of-phase precession defines the OM, labeled $\delta\boldsymbol{m}_-$. These are expressed as $\delta\boldsymbol{m}_\pm = \delta\boldsymbol{m}_1 \pm C_{2z'}\delta\boldsymbol{m}_2$, where $C_{2z'}$ denotes a 2-fold (180°) rotation about the axis $\boldsymbol{z'}$ that is parallel to the equilibrium vector sum $\boldsymbol{m}_1^{eq} + \boldsymbol{m}_2^{eq}$ [41]. Similarly, for the RM-LM coupling regime, the RM and LM chiral modes correspond to right- and left-circular precessions, respectively. They are defined as $\delta\boldsymbol{m}_{j\pm} = \delta\boldsymbol{m}_{j\varphi} \pm i\delta\boldsymbol{m}_{j\theta}, j = 1{,}2$, where $\delta\boldsymbol{m}_{j\varphi}$ and $\delta\boldsymbol{m}_{j\theta}$ are the orthogonal in-plane components of the magnetization precession. Using these definitions, the eigenvalue equations for the uncoupled modes are readily derived (see

SM [40] for details), yielding the frequency-dispersion formulas that fit the experimental data (solid curves in Figs. 1(a–h)). Setting the off-diagonal blocks of the dynamical matrix to zero gives the corresponding dispersions of the pure RM, LM, OM, and AM modes, plotted as dashed lines in Figs. 1(a–h).

**Quantum model:** While the classical LL model successfully describes the mode evolution, it proves inadequate for precisely quantifying coupling strengths in asymmetric systems [22,24]. This limitation necessitates a quantum model beyond the RWA. Similar to the approach developed for SAF and AFM [4,30], we decompose the Hamiltonian into uncoupled and interacting components (see Section III of SM [40]):

$$\hat{\mathcal{H}}_{all} = \hat{\mathcal{H}}_{pure} + \hat{\mathcal{H}}_{couple} \,. \quad (1)$$

For the AM-OM coupling regime, the Hamiltonian takes the form:

$$\hat{\mathcal{H}}_{pure} = \hbar\omega_+ \left(\hat{a}_+^\dagger \hat{a}_+ + \frac{1}{2}\right) + \hbar\omega_- \left(\hat{a}_-^\dagger \hat{a}_- + \frac{1}{2}\right),$$
$$\hat{\mathcal{H}}_{couple} = i\hbar g_1 \left(\hat{a}_+ \hat{a}_-^\dagger - \hat{a}_+^\dagger \hat{a}_-\right) + i\hbar g_2 \left(\hat{a}_+^\dagger \hat{a}_-^\dagger - \hat{a}_+ \hat{a}_-\right), \quad (2)$$

where $\omega_+$ and $\omega_-$ correspond to the frequencies of pure AM and OM, respectively, and $\hat{a}_+$ ($\hat{a}_+^\dagger$) and $\hat{a}_-$ ($\hat{a}_-^\dagger$) are their associated bosonic annihilation (creation) operators. The co-rotating coupling term $g_1$ represents the coherent magnon exchange leading to AM-OM hybridization, while the counter-rotating coupling $g_2$ accounts for non-energy-conserving processes that create or annihilate magnon pairs, generating two-mode squeezed states. Crucially, our experimental fits reveal $g_2 > g_1$ [Fig. 1(i)], demonstrating the breakdown of the RWA and the necessity of including counter-rotating terms in USC regimes ($g/f_g \geq 0.1$). This invalidates conventional weak-coupling approaches that neglect such quantum fluctuations and their associated phenomena like magnon squeezing.

For the RM-LM coupling case, the Hamiltonian structure differs significantly for chiral modes:

$$\hat{\mathcal{H}}_{pure} = \hbar\omega_+ \hat{a}_+^\dagger \hat{a}_+ + \hbar\omega_- \hat{a}_-^\dagger \hat{a}_- \,,$$
$$\hat{\mathcal{H}}_{couple} = i\hbar g_1 \left(\hat{a}_+ \hat{a}_-^\dagger - \hat{a}_+^\dagger \hat{a}_-\right) + \hbar \left[g_{3+} \left(\hat{a}_+^2 + \hat{a}_+^{\dagger\,2}\right) - g_{3-} \left(\hat{a}_-^2 + \hat{a}_-^{\dagger\,2}\right)\right], \quad (3)$$

In this case, the self-squared terms ($g_{3\pm}$) replace the two-mode coupling $g_2$, enabling single-mode squeezing through RM-RM and LM-LM pair processes. The net coupling $g_3 = g_{3+} + g_{3-}$ becomes comparable to $g_1$ [Fig. 1(j)], with the quantum-derived $g$ consistently yielding half the classical minimal gap (spherical symbols in Figs. 1(i-k)), validating our approach.

The coupling strengths exhibit distinct behaviors across AM-OM and RM-LM coupling regimes. In the RM-LM phase ($AFM_c$ state under $H \parallel z$), the couplings $g_1$ and $g_3$ show weak field dependence below the spin-flop transition, as the $Cr^{2+}$ moments remain nearly antiparallel along the *c*-axis with minimal canting. In the AM-OM regime (canted $AFM_b$ state under $H \parallel z$), $g_1$ and $g_2$ vary strongly with $H$ as a consequence of the substantial spin reorientation toward the field direction after the spin-flop transition. Both regimes display enhanced coupling at lower temperatures, consistent with the enhanced $H_A$ and $H_E$ [Fig. S7(b) [40]], as predicted theoretically in Ref. [42], which relates the coupling terms to the increasing anisotropy ($H_A$) and exchange ($H_E$) fields. The normalized coupling ratios $g/f_g$ in Fig. 1(k) are overall greater than 0.1, confirming that both interaction types operate in the USC regime. In this regime, the strengths of the quantum fluctuation terms ($g_2$, $g_3$) become comparable to the coherent hybridization term ($g_1$), indicating a significant deviation from the RWA. This comprehensive quantum model bridges the gap between the microscopic Hamiltonian and the macroscopic spectroscopic observables in this A-type layered AFM.

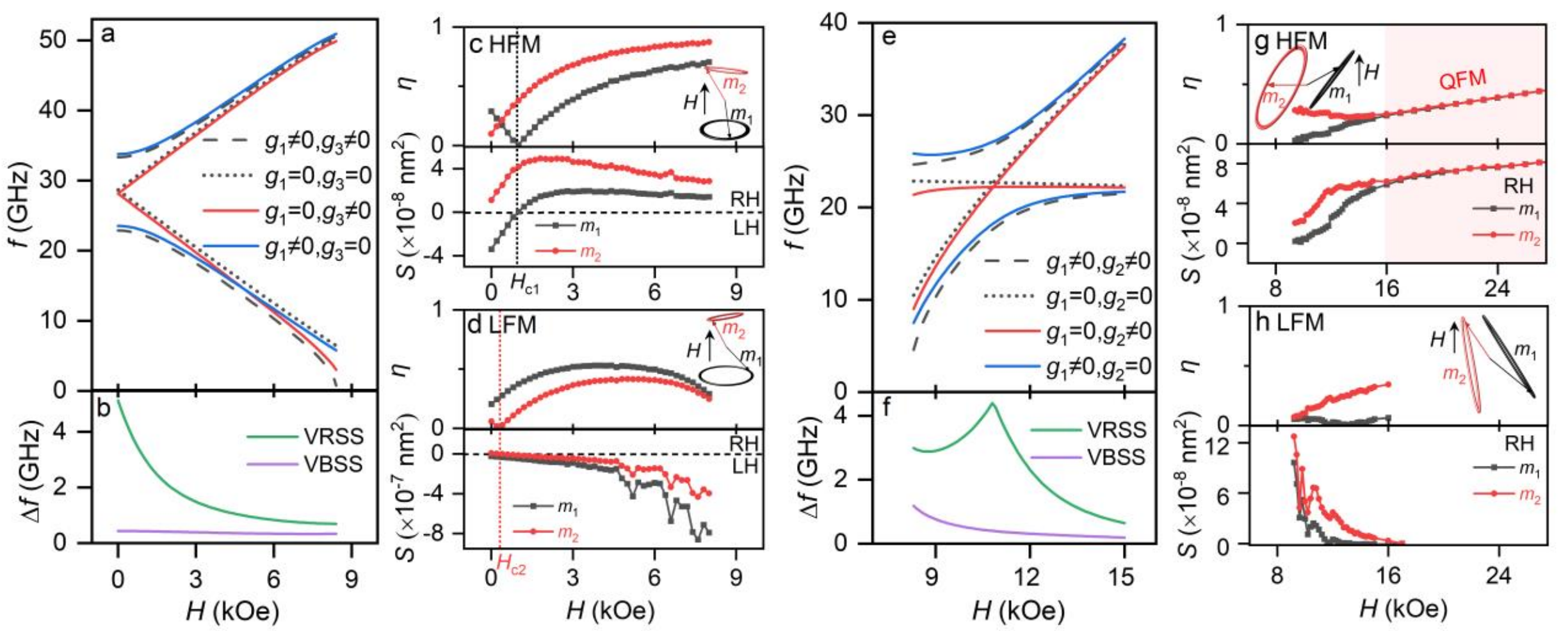

**Figure 2.** Field-dependent VRSS, VBSS, and spin precession trajectory characteristics. (a) The experimental fitting (dashed lines) and three theoretically calculated frequency dispersion curves under an out-of-plane field at 7 K in the low-field RM-LM regime. (b) Calculated field dependence of VRSS (blue line) and VBSS (pink line), as defined in the main text. (c, d) The ellipticity $\eta$ describing trajectory polarization and the directed orbit area $S$ of $m_1$ (red) and $m_2$ (black) obtained from micromagnetic simulation for the coupled high-frequency mode (HFM) (c) and low-frequency mode (LFM) (d). The absolute value of $S$ indicates excitation strength (spin-wave intensity), while its sign denotes chirality: positive for right-handed (RH) and negative for left-

handed (LH). The vertical dotted lines mark the critical field $H_c$ for chirality reversal. (e–h) Same as (a–d) for the high-field AM-OM regime. The red background in (g) corresponds to the quasi-ferromagnetic resonance mode (QFM) region. Inset: typical schematic precession orbits of $m_1$ and $m_2$ at zero field (c, d) and at $H$ = 9.4 kOe (g, h).

**Squeezed single- and two-mode magnon states:** To elucidate the distinct roles of $g_1$, $g_2$, and $g_3$ in mode hybridization, we analyze the associated frequency shifts: the vacuum Rabi splitting shift (VRSS) and the vacuum Bloch-Siegert shift (VBSS). The VRSS, originating from the co-rotating term $g_1$, is defined as the difference between the high-frequency mode $\omega_+$ ($g_1 \neq 0$, $g_2$ ($g_3$) = 0) with only $g_1$ active and the higher bare mode frequency $\omega_{\max}$ ($g_1$ = 0, $g_2$ ($g_3$) = 0) with all couplings turned off. The VBSS, a hallmark of physics beyond the RWA, arises from the counter-rotating terms and is given by the difference between $\omega_+$ ($g_1 \neq 0$, $g_2$ ($g_3$) = 0) with only $g_1$ active and $\omega_+$ ($g_1 \neq 0$, $g_2$ ($g_3$) $\neq$ 0) in the full model (see SM Section III [40]). As shown in Figs. 2(a) and (e), $g_1$ dominates the mode anticrossing, with VRSS maximized at the minimal frequency gap. Figures 2(b) and (f) further show that the field-dependent VBSS closely follows the dominant counter-rotating term in each regime, tracking $g_3$ in the RM-LM regime and $g_2$ in the AM-OM regime. This directly links the VBSS to the single-mode squeezing ($g_3$) and two-mode squeezing ($g_2$) channels, respectively.

To visualize the chirality and polarization of these hybrid quantum states, we perform micromagnetic simulations, which well reproduce the spectroscopic results [Fig. S5 [40]]. In the low-field RM-LM regime, the high-frequency mode (HFM) exhibits a chirality reversal for the sublattice moment $m_1$ (from LH to RH) at a critical field $H_{c1}$ = 1 kOe, while $m_2$ remains RH [Fig. 2(c)]. Conversely, for the low-frequency mode (LFM), $m_2$ reverses chirality (RH to LH) at a lower field $H_{c2}$ = 0.2 kOe, while $m_1$ stays LH [Fig. 2(d)]. These complementary reversals indicate strong RM-LM intermode coupling. The LFM also shows a larger precession amplitude and lower ellipticity than the HFM. In contrast, in the high-field AM-OM regime, both sublattices precess with identical RH chirality [Figs. 2(g, h)]. As the field increases, spectral weight transfers from the LFM to the HFM, with the LFM amplitude vanishing near the spin-flop transition (~16 kOe), where the HFM becomes a quasi-ferromagnetic mode (QFM). Precise control over chirality and polarization via the bias field H provides a practical route to manipulate the underlying quantum states, such as the squeezed magnon vacuum, using microwave excitations.

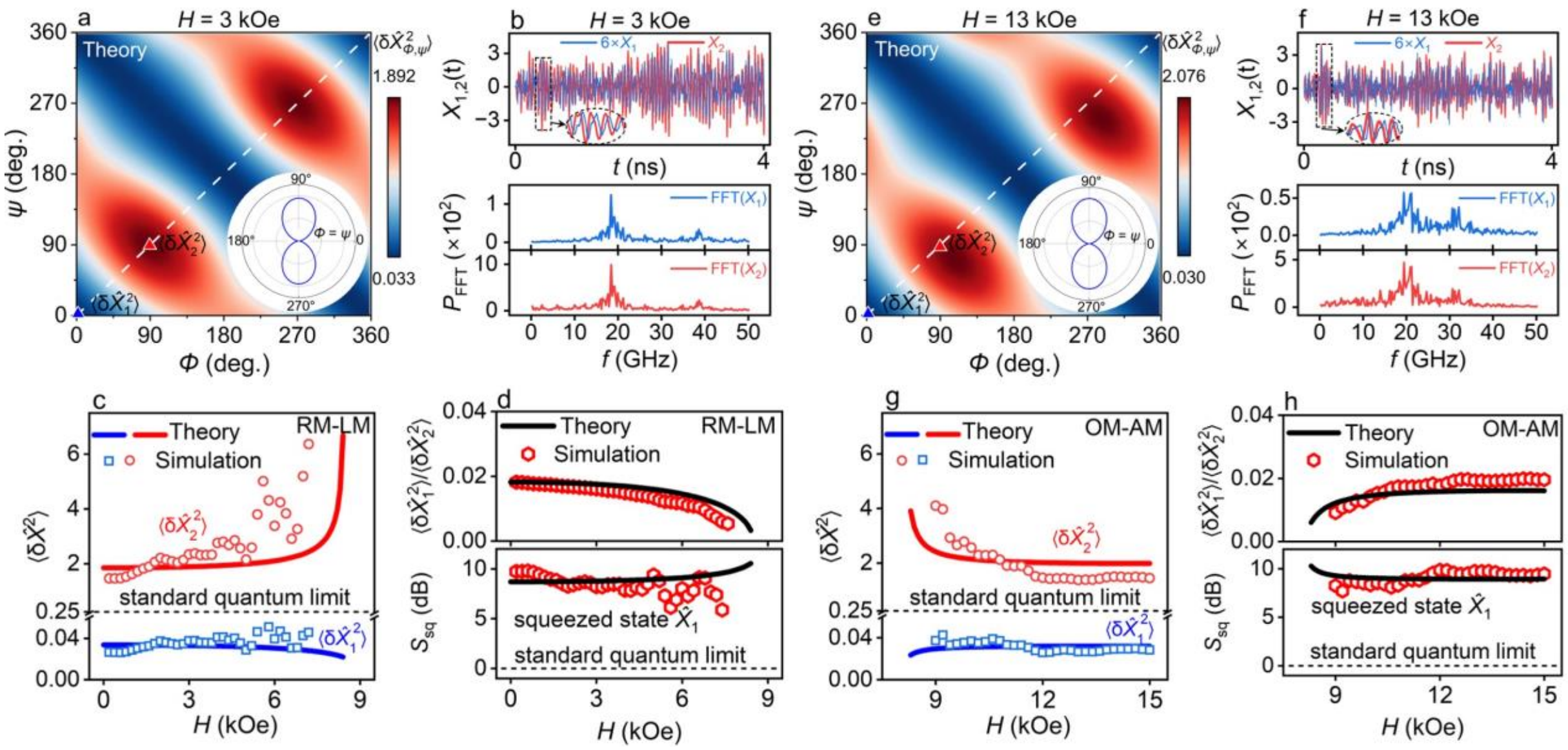

**Figure 3.** Theoretical and simulated quadrature fluctuations and squeezing ratios. (a) Theoretical color map of quadrature fluctuations as functions of phase angles $\Phi$ and $\psi$ at $H$ = 3 kOe. White dashed lines indicate the cut along $\Phi = \psi$; its polar plot is shown in the inset. Blue and red triangles mark the fluctuation $\langle\delta\hat{X}_1^2\rangle$ at ($\Phi$, $\psi$) = (0°, 0°) and $\langle\delta\hat{X}_2^2\rangle$ at ($\Phi$, $\psi$) = (90°, 90°), respectively. (b) Simulated $X_1$ and $X_2$ quadrature time traces over 4 ns and corresponding Fourier magnitudes $P_{\mathrm{FFT}}$ at the same field $H$ = 3 kOe. (c) Comparison between theoretical and simulated $\langle\delta\hat{X}_1^2\rangle$ and $\langle\delta\hat{X}_2^2\rangle$ against the SQL (0.25) in the low-field RM-LM regime. (d) Corresponding squeezing ratios $\langle\delta\hat{X}_1^2\rangle/\langle\delta\hat{X}_2^2\rangle$ and squeezing factors $S_{\mathrm{sq}}$ (dB) = $-10\log_{10}(\langle\delta\hat{X}_1^2\rangle/0.25)$ in the RM-LM regime. (e–h) Same as (a–d) but for $H$ = 13 kOe (e, f) and the high-field AM-OM regime (g, h).

**Quadrature fluctuations and squeezing ratios:** To quantify the squeezing of the two-sublattice magnon modes, we introduce a general quadrature operator constructed from the creation and annihilation operators for the two sublattices, $a_1$ and $a_2$, with two independent phase angles $\Phi$ and $\psi$ [33]:

$$\hat{X}_{\phi,\psi} = \frac{1}{\sqrt{8}}\left[\hat{a}_1 e^{-i\phi} + \hat{a}_1^{\dagger} e^{i\phi} + \hat{a}_2 e^{-i\psi} + \hat{a}_2^{\dagger} e^{i\psi}\right]. \quad (4)$$

In the entangled ground state $|0_{\mathrm{H}}, 0_{\mathrm{L}}\rangle$ of the hybridized magnon modes, we calculate the quadrature fluctuation $\langle\delta\hat{X}_{\phi,\psi}^2\rangle$, which directly probes the redistribution of quantum noise induced by the USC. The representative results are shown in Fig. 3(a) for the low-field RM-LM regime at $H$ = 3 kOe and in Fig. 3(e) for the AM-OM regime at $H$ = 13 kOe. These fluctuations satisfy the uncertainty relation $\langle \delta\hat{X}^2_{\phi,\psi}\rangle \cdot \left\langle \delta\hat{X}^2_{\phi+\frac{\pi}{2},\psi+\frac{\pi}{2}}\right\rangle \geq \frac{1}{16}$. For an uncoupled coherent state, the variance for any quadrature saturates at the SQL of 0.25. Specifically, $\Phi = \psi = 0$ yields the minimum fluctuation $\langle \delta\hat{X}^2_1\rangle$, while $\Phi = \psi = 90°$ gives the maximum fluctuation $\langle \delta\hat{X}^2_2\rangle$. These two conjugate quadratures define the principal squeezing axes: the variance of one is reduced below the SQL, while that of the orthogonal one is enhanced, with their product remaining bounded by the uncertainty principle.

Through the Holstein-Primakoff transformation, these quadratures can be directly mapped onto experimentally accessible magnetization components [4, 28, 33],

$$\hat{X}_1 = \frac{1}{\sqrt{2}\sqrt{2\gamma_0\hbar M_s}}\left[\hat{M}_{1\varphi} + \hat{M}_{2\varphi}\right],$$

$$\hat{X}_2 = \frac{-1}{\sqrt{2}\sqrt{2\gamma_0\hbar M_s}}\left[\hat{M}_{1\theta} + \hat{M}_{2\theta}\right], (5)$$

where $\theta$ and $\varphi$ denote the out-of-plane and in-plane components, respectively, of the spin precession with respect to the equilibrium $M_1$-$M_2$ plane of the two sublattices. In the low-field RM-LM configuration, $\boldsymbol{X_1}$ corresponds to the in-plane component of the total spin, while $\boldsymbol{X_2}$ represents the out-of-plane component of the Néel order. In contrast, for the high-field AM-OM configuration, $\boldsymbol{X_1}$ and $\boldsymbol{X_2}$ represent the in-plane and out-of-plane components of the Néel order, respectively (see Fig. S4 of the SM [40]).

We further demonstrate the fluctuations of the two magnetization quadratures, $\boldsymbol{X_1}$ and $\boldsymbol{X_2}$, using micromagnetic simulations that incorporate classical thermal noise [43]. Representative time traces for the RM-LM regime at 3 kOe and the AM-OM regime at 13 kOe are shown in Figs. 3(b) and (f), respectively. The traces of $\boldsymbol{X_1}$ and $\boldsymbol{X_2}$ exhibit a 90° phase shift, confirming their orthogonality. By integrating the power spectral density around the eigenmode frequencies, we obtain the fluctuations $\langle \delta\hat{X}^2_1\rangle$ and $\langle \delta\hat{X}^2_2\rangle$ at each field. Their values and ratios are summarized in Figs. 3(c, d) and 3(g, h). The quantitative agreement between the simulated fluctuations and the predictions of our quantum model validates our entire protocol: from the microscopic Hamiltonian, to the mapping onto physical observables [Eq. (5)], and finally to the extraction of quantum properties from classical spin dynamics. To quantify the degree of noise suppression, we adopt the standard metric from quantum optics: the squeezing factor $S_{sq}$ (dB) = $-10\log_{10}(\langle \delta\hat{X}^2_1\rangle/0.25)$, where 0.25 is the variance of the unsqueezed (coherent) state. A value above 0 dB indicates quadrature squeezing, with larger values corresponding to stronger noise reduction. In our simulations, $S_{sq}$ reaches up to ~9.7 dB. This substantial predicted squeezing highlights the potential of $CrPS_4$ as a novel solid-state platform for generating and manipulating non-classical magnon states, complementary to established microwave squeezing platforms such as Josephson/kinetic inductance parametric amplifiers [44–46]. Future experiments can measure the time-dependent transverse magnetization components $M_{i\theta}(t)$ and $M_{i\varphi}(t)$ via, e.g., time-resolved magneto-optical Kerr effect or nitrogen-vacancy center magnetometry.

Our work establishes a direct and practical protocol for detecting and quantifying quantum squeezing in AFMs, covering both the AM-OM regime and the previously inaccessible RM-LM regime. By mapping abstract quadrature operators onto experimentally measurable magnetization dynamics, and validating the approach with micromagnetic simulations, we provide an actionable framework for identifying magnon squeezing in real materials. This bridges quantum-optical metrology and spintronics, opening a route toward exploiting squeezed magnons for quantum-enhanced sensing, noise-resilient signal transduction, and hybrid quantum information platforms.

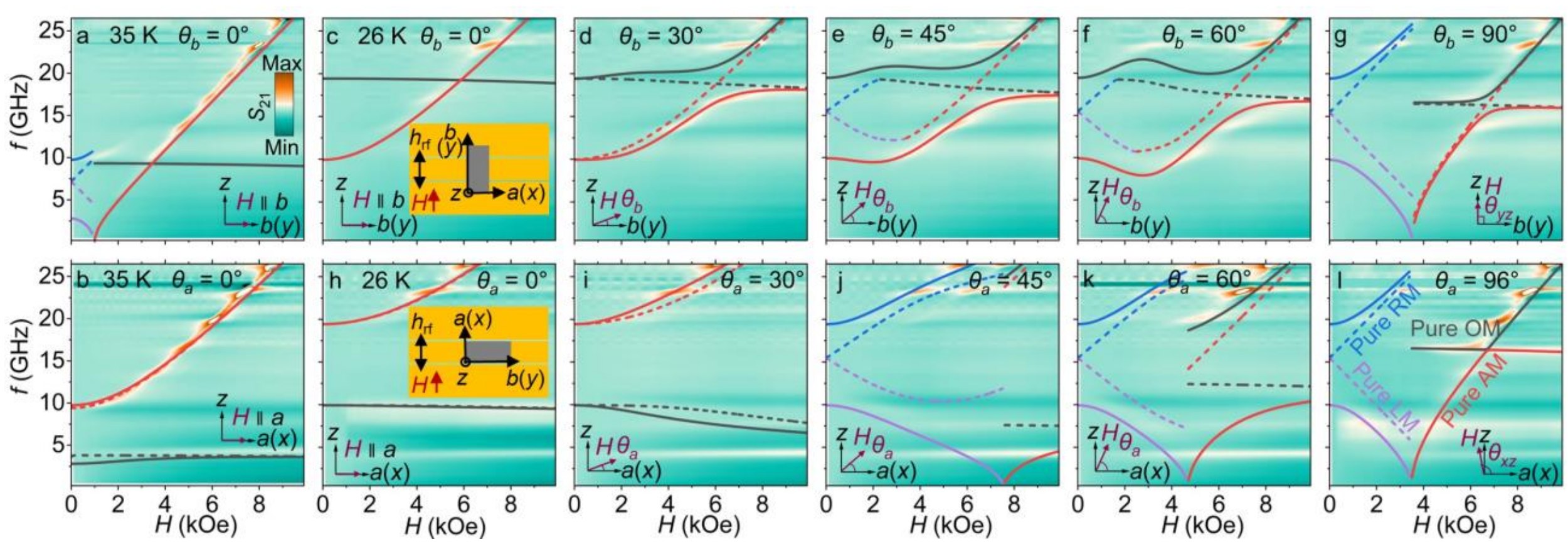

**Figure 4.** VNA-AFMR $S_{21}$ spectra and theoretical dispersion curves under different field angles $\theta_b$ and $\theta_a$. (a, b) Color plots of the experimental AFMR spectra with $H$ along the $b$-axis (a), the $a$-axis (b) at 35 K. (c–l) Color plots of the experimental AFMR spectra with the labeled angles $\theta_b$ (c–g) and $\theta_a$ (h–l) at 26 K. Dashed and solid lines represent theoretically calculated pure modes without mode coupling and hybridized modes, respectively. Insets: schematics of the field angles $\theta_b$, $\theta_a$ and the experimental setup.

**Angular dependence of hybridization modes:** The orientation of the external magnetic field provides an additional dimension for controlling magnon-magnon coupling in $CrPS_4$, complementing temperature and field magnitude. This angular dependence allows us to selectively activate, suppress, or hybridize the RM-LM and AM-OM coupling regimes. At 35 K ($AFM_b$ phase), applying $H$ along the $b$-axis breaks the dynamical equivalence of the two sublattices, activates the off-diagonal terms in the dynamical matrix and leads to RM-LM hybridization at low fields ($H < H_{sf}$) [Fig. 4(a)]. At high fields ($H \gg H_{sf}$), the two sublattice moments become nearly collinear, and the AM and OM modes effectively decouple and cross [Fig. 4(a)]. This symmetry-governed picture is further validated at 26 K ($AFM_c$ phase). A pure, decoupled AM and OM spectrum emerges at specific angles (e.g., $\theta_b$ = 0°, $\theta_a$ = 96°) where the effective internal field drives the sublattice moments symmetrically with $H$ [Figs. 4(c, l)]. At other field orientations [Figs. 4(d–g, h–k)], this symmetry is broken, and the system enters a regime of complex hybridization in which both RM-LM and AM-OM coupling channels are engaged simultaneously. The full complexity of these spectra, arising from the interplay of all coupling terms ($g_1$, $g_2$, $g_3$), is well captured by our complete quantum model and micromagnetic simulations, as shown by the agreement between theory and data in Fig. 4 and Fig. S5 [40], respectively. The coupling strength, approximately indicated as the frequency gap, is maximized when the field angle $\theta_a$ or $\theta_b$ approaches 45°, where the external perturbation most efficiently mixes the original eigenmodes. In brief, magnetic field orientation acts as a continuous tuning knob that can switch coupling on or off via symmetry control and selectively enhance different interaction channels. Together with temperature and field magnitude, it enables control of quantum magnonic states in this low-symmetry vdW magnet.

**In summary**, we established a comprehensive quantum framework for magnon-magnon coupling in the triaxial vdW AFM $CrPS_4$. By integrating AFMR spectroscopy with a microscopic quantum model beyond the RWA, we isolated three distinct interaction channels: the co-rotating term ($g_1$, hybridizing), the counter-rotating term ($g_2$, two-mode squeezing), and, most importantly, the self-squared term ($g_3$, single-mode squeezing). The system reaches the USC regime with a normalized coupling ratio $g/f_g$~0.41, where these terms collectively reshape the quantum ground state, generating a strongly squeezed magnon vacuum with a squeezing factor up to ~9.7 dB. By linking the quantum model to micromagnetic simulations, we demonstrated a direct protocol to extract quadrature fluctuations and squeezing ratios from measurable magnetization dynamics. The coupling strength, chirality, polarization, and the resulting quantum states are tunable by temperature and magnetic field, highlighting the unique level of control afforded by low-symmetry, low-damping vdW magnets. Our work opens a route toward functional quantum devices, with switchable $g_2/g_3$ enabling tunable nonclassical spin states, quantum-enhanced sensing, and hybrid quantum functionalities.

## Acknowledgments

We acknowledge the financial support from the National Key Research and Development Program of China (Grant No. 2023YFA1406603, 2024YFA1408801) and the National Natural Science Foundation of China (Grant No. 12574125).

# Supplemental Material

## Tunable Ultrastrong Magnon-Magnon Coupling and Squeezed States in a Triaxial Van der Waals Antiferromagnet

Xiao Xiao[1], Zishuang Li[1], Zui Tao[1], Lina Chen[2,3], Ping Liu[2], Shuai Zhang[1], Haozhe Wang[1], Yong Pu[2] and Ronghua Liu[1,4]*

[1] National Key Laboratory of Spintronics and School of Physics, Nanjing University, Nanjing 210093, China

[2] School of Science, Nanjing University of Posts and Telecommunications, Nanjing 210023, China

[3] Laboratory of Solid State Microstructures, Nanjing University, Nanjing 210093, China

[4] State Key Laboratory of Spintronics, Nanjing University, Suzhou 215163, China

*Corresponding Email: rhliu@nju.edu.cn

Supplemental Material

## Section I. Magnetic phase diagrams

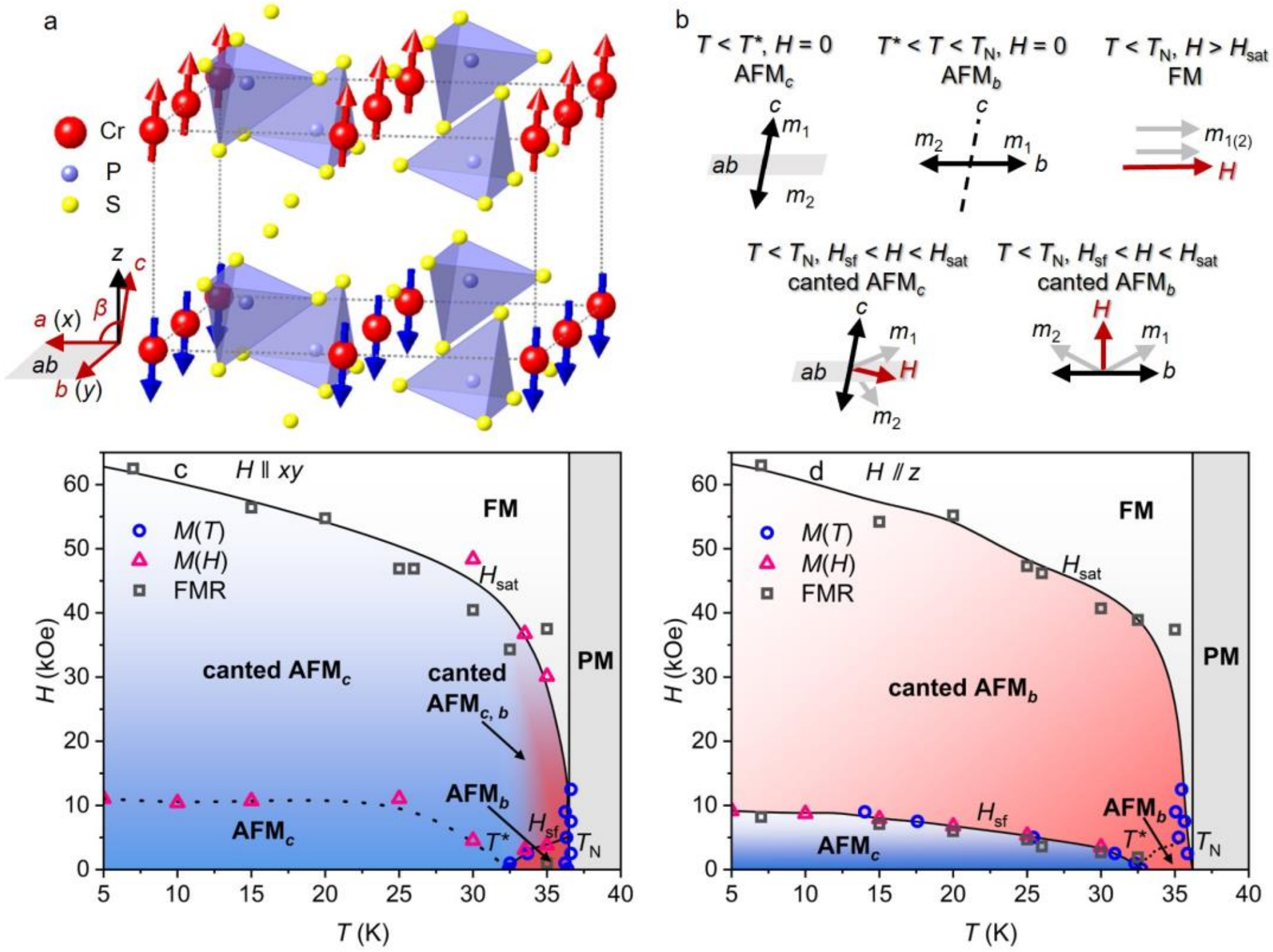

**Figure S1.** Magnetic structure and phase diagrams of $CrPS_4$. (a) Crystal and magnetic structure of $CrPS_4$ at $T < T^*$. $Cr^{2+}$ spin magnetic moments are along the easy axis $c$ of the orthorhombic unit cell. (b) Schematic illustrations of the spin structures in each magnetic phase. $m_1$ and $m_2$ are the magnetization directions of the two AFM sublattices. $AFM_{c(b)}$ and canted-$AFM_{c(b)}$ are the ground states with spins aligned antiparallel and canted antiparallel along the $c$($b$)-axis, respectively. $T^*/T_N$ is the spin-reorientation/Néel temperature, and $H_{sf}/H_{sat}$ is the spin-flop/saturation field. (c–d) $H$-$T$ phase diagrams constructed based on $M(T)$, $M(H)$, and FMR experiment data for (c) $H \parallel xy$ and (d) $H \parallel z$.

$CrPS_4$ is a quasi-2D monoclinic crystal structure with triaxial magnetic anisotropies and exhibits A-type antiferromagnetism, with $Cr^{2+}$ spins aligned ferromagnetically within each layer and antiferromagnetically coupled along the monoclinic easy $c$-axis of the orthorhombic unit cell ($\beta$ = 91.88° between $a$ and $c$-axis, with $b$ perpendicular to $a$ and $c$ axes) [Fig. S1(a)]. $CrPS_4$ exhibits distinct temperature- and field-dependent static magnetic configurations due to its layered structure and strong spin-orbit coupling, as illustrated schematically in Fig. S1(b). Using SQUID measurements, we determined the field-temperature magnetic phase diagrams for $H \parallel xy$ plane and $H \parallel z$, shown in Figs. S1(c) and (d). The phase boundaries are identified by the extrema of $d(\chi T)/dT$ from *M*-*T* curves and $dM/dH$ from *M*-*H* curves, following magnetic phase transition theory [1–4]. We determine the Néel temperature $T_N$ to be 36.5 K from the susceptibility data. When the temperature cools down to $T^*$ = 32.5 K, it undergoes a spin reorientation from the ground state $AFM_b$ to $AFM_c$ due to the maximum magnetic easy axis switching from the *b*-axis to the *c*-axis. This Morin-type transition reflects the spin reorientation that occurs under cooling, also reported in α-$Fe_2O_3$ [5]. By combining field- and temperature-dependent AFMR [Fig. 1 in main text] and magnetic susceptibility results, the spin-flop field $H_{sf}$ and saturation field $H_{sat}$ are summarized in the magnetic phase diagrams. To explore magnetodynamics across different phases, we further measure field- and angular-dependent AFMR dispersions at different temperatures and crystallographic planes.

**Section II. Theoretical derivation of the classical LL model**

The triaxial magnetic anisotropy energy of $CrPS_4$ can be represented, including an out-of-plane monoclinic easy-axis (*c*-axis) anisotropy and two in-plane orthorhombic easy-axis (*a*- and *b*-axis) anisotropies [6]

$$E_{Anis} = \frac{K_1'}{2}\left({\alpha_a}^2 + {\beta_a}^2\right) + \frac{K_2'}{2}\left({\alpha_b}^2 + {\beta_b}^2\right) + \frac{K_3'}{2}\left({\alpha_c}^2 + {\beta_c}^2\right)$$

$$= \left[\frac{K_1'}{2} + \frac{K_3'}{2}(cos^2\beta - sin^2\beta)\right]\left({\alpha_1}^2 + {\beta_1}^2\right) + \left[\frac{K_2'}{2} - \frac{K_3'}{2}sin^2\beta\right]\left({\alpha_2}^2 + {\beta_2}^2\right)$$

$$+ K_3' sin\beta cos\beta(\alpha_1\alpha_3 + \beta_1\beta_3)$$

$$= \frac{K_1}{2}\left({\alpha_1}^2 + {\beta_1}^2\right) + \frac{K_2}{2}\left({\alpha_2}^2 + {\beta_2}^2\right) + K_m(\alpha_1\alpha_3 + \beta_1\beta_3), \quad \text{(S1)}$$

where $\alpha_i$ and $\beta_i$ ($i$ = *a*, *b*, and *c*) represent the cosines of $m_1$ and $m_2$ relative to the three crystal axes *a*, *b*, and *c*, respectively; while $\alpha_j$ and $\beta_j$ ($j$ = 1, 2, and 3) represent the cosines of $m_1$ and $m_2$ relative to the coordinate axes $x$, $y$, and $z$. The transformation between the two coordinate systems is given by $\alpha_a = \alpha_1$, $\alpha_b = \alpha_2$, $\alpha_c = \alpha_1\cos\beta + \alpha_3\sin\beta$; $\beta_a = \beta_1$, $\beta_b = \beta_2$, $\beta_c = \beta_1\cos\beta + \beta_3\sin\beta$.

The expression for the total energy $E$ is formulated as follows:

$$E = E_{Zeeman} + E_{Demag} + E_{Anis} + E_{Exchange}$$

$$E_{Zeeman} = \sum_{i=1}^{2}[-\mu_0 M_s \boldsymbol{H}\cdot\boldsymbol{m_i}]$$

$$E_{Demag} = \sum_{i=1}^{2}\left[\frac{1}{2}\mu_0 {M_s}^2(\boldsymbol{m_i}\cdot\hat{\boldsymbol{z}})^2\right]$$

$$E_{Anis} = \sum_{i=1}^{2}\left[\mu_0 M_s\left[\frac{H_{A1}}{2}(\boldsymbol{m_i}\cdot\hat{\boldsymbol{x}})^2 + \frac{H_{A2}}{2}(\boldsymbol{m_i}\cdot\hat{\boldsymbol{y}})^2 + H_{Am}(\boldsymbol{m_i}\cdot\hat{\boldsymbol{x}})(\boldsymbol{m_i}\cdot\hat{\boldsymbol{z}})\right]\right]$$

$$E_{Exchange} = -\mu_0 M_s H_E(\boldsymbol{m_1}\cdot\boldsymbol{m_2}), \qquad \boldsymbol{i} = 1, 2\,, \qquad \text{(S2)}$$

where the terms represent Zeeman energy, demagnetization energy, anisotropy energy (effective anisotropy fields $H_{A1}$, $H_{A2}$, $\frac{H_{Am}}{sin\beta cos\beta}$ for the crystal axes $a$, $b$, and $c$), and interlayer RKKY exchange energy (exchange field $H_E$), respectively.

Considering the general situation, the angles between the external magnetic field $H$ and the $x$- and $z$-axis are defined as $\theta$ and $\varphi$, respectively. The orientations of the two magnetic moments, $m_1$ and $m_2$, are defined by the polar ($\theta_1$, $\theta_2$) and azimuthal ($\varphi_1$, $\varphi_2$) angles. The equilibrium configuration of $m_{1,2}$, determined by minimizing the total energy $E$, can thus be expressed in terms of these angular parameters. By neglecting the dissipative term, the magnetic dynamics of this triaxial magnetic anisotropic antiferromagnet can be described by the Landau-Lifshitz (LL) equations [7]

$$\frac{d\boldsymbol{m_i}}{dt} = -\gamma\boldsymbol{m_i}\times\boldsymbol{H_{eff,i}}, \qquad \boldsymbol{H_{eff,i}} = -\frac{1}{\mu_0 M_s}\nabla_{\boldsymbol{m_i}}\boldsymbol{E}, \qquad \boldsymbol{i} = 1, 2\,. \quad \text{(S3)}$$

To linearize Eq. (S3), we further decompose the precessing magnetization into the direct and alternating components

$$\boldsymbol{m_i} = \boldsymbol{m_i^{eq}} + \delta \boldsymbol{m_i} \boldsymbol{e^{i\omega t}}, \qquad \boldsymbol{i} = 1, 2\,. \quad \text{(S4)}$$

In the Cartesian coordinate system (*x*, *y*, *z*), to solve for the resonance frequencies by substituting Eq. (S4) into Eq. (S3), we obtain the following eigenvalue equation in matrix form:

$$-\frac{i\omega}{\gamma}\begin{pmatrix}\delta \boldsymbol{m_{1x}}\\ \delta \boldsymbol{m_{1y}}\\ \delta \boldsymbol{m_{1z}}\\ \delta \boldsymbol{m_{2x}}\\ \delta \boldsymbol{m_{2y}}\\ \delta \boldsymbol{m_{2z}}\end{pmatrix} = A\begin{pmatrix}\delta \boldsymbol{m_{1x}}\\ \delta \boldsymbol{m_{1y}}\\ \delta \boldsymbol{m_{1z}}\\ \delta \boldsymbol{m_{2x}}\\ \delta \boldsymbol{m_{2y}}\\ \delta \boldsymbol{m_{2z}}\end{pmatrix}, \qquad A = \begin{bmatrix} B & C \\ C & B \end{bmatrix}, \qquad \text{(S5)}$$

where B and C are 3 × 3 matrices derived from the terms $\boldsymbol{m}_1 \times \boldsymbol{H}_{\mathrm{eff},1}$ and $\boldsymbol{m}_2 \times \boldsymbol{H}_{\mathrm{eff},2}$, respectively. Their explicit form is given by

$$A = \begin{bmatrix} B_{11} & B_{12} & B_{13} & 0 & C_{12} & C_{13} \\ B_{21} & 0 & B_{23} & C_{21} & 0 & C_{23} \\ B_{31} & B_{32} & B_{33} & C_{31} & C_{32} & 0 \\ 0 & C_{12} & C_{13} & B_{11} & B_{12} & B_{13} \\ C_{21} & 0 & C_{23} & B_{21} & 0 & B_{23} \\ C_{31} & C_{32} & 0 & B_{31} & B_{32} & B_{33} \end{bmatrix}, \qquad \text{(S6)}$$

where the elements are defined as follows:

$$B_{11} = -H_{Am} \sin\theta_1\, sin\varphi_1$$

$$B_{12} = (H_{A2} - M_s)\cos\theta_1 + H_E \cos\theta_2 - H_{Am} \sin\theta_1\, cos\varphi_1 + Hcos\theta$$

$$B_{13} = (H_{A2} - M_s)\sin\theta_1\, sin\varphi_1 - H_E \sin\theta_2 sin\varphi_2 - Hsin\theta sin\varphi$$

$$B_{21} = (M_s - H_{A1})\cos\theta_1 - H_E\cos\theta_2 + 2H_{Am}\sin\theta_1\, cos\varphi_1 - Hcos\theta$$

$$B_{23} = (M_s - H_{A1})\sin\theta_1\, cos\varphi_1 + H_E \sin\theta_2\, cos\varphi_2 - 2H_{Am}\cos\theta_1 + Hsin\theta sin\varphi$$

$$B_{31} = (H_{A1} - H_{A2})\sin\theta_1\, sin\varphi_1 + H_E\sin\theta_2\, sin\varphi_2 + Hsin\theta sin\varphi$$

$$B_{32} = (H_{A1} - H_{A2})\sin\theta_1\, cos\varphi_1 - H_E\sin\theta_2\, cos\varphi_2 + H_{Am}\cos\theta_1 - Hsin\theta cos\varphi$$

$$B_{33} = H_{Am}\sin\theta_1\, sin\varphi_1$$

$$C_{12} = -H_E\cos\theta_1$$

$$C_{13} = H_E\sin\theta_1\, sin\varphi_1$$

$$C_{21} = H_E \cos\theta_1$$

$$C_{23} = -H_E \sin\theta_1\, cos\varphi_1$$

$$C_{31} = -H_E \sin\theta_1\, sin\varphi_1$$

$$C_{32} = H_E \sin\theta_1\, cos\varphi_1.$$

To facilitate the discussion of the spin dynamics of the acoustic mode (AM) and optical mode (OM) coupling regime, we first transform the original global coordinate system ($x$, $y$, $z$) to a local coordinate system ($\varphi$, $\theta$, $m$), in which the basis vector $\boldsymbol{e_m}$ is aligned with the direct component of the magnetic moment $m_1$. This transformation is performed via a matrix $P$:

$$\begin{pmatrix} \boldsymbol{e_\varphi} \\ \boldsymbol{e_\theta} \\ \boldsymbol{e_m} \end{pmatrix} = P \begin{pmatrix} \boldsymbol{e_x} \\ \boldsymbol{e_y} \\ \boldsymbol{e_z} \end{pmatrix}, \qquad \text{(S7)}$$

where $\boldsymbol{e_x}$, $\boldsymbol{e_y}$, $\boldsymbol{e_z}$ and $\boldsymbol{e_\varphi}$, $\boldsymbol{e_\theta}$, $\boldsymbol{e_m}$ are the bases of the global and local coordinate systems, respectively. The local coordinate system is constructed with respect to the magnetic moment $m_1$ itself, and the local basis vectors are defined as

$$\boldsymbol{e_m} = \frac{\boldsymbol{m_1^{eq}}}{|\boldsymbol{m_1^{eq}}|}, \qquad \boldsymbol{e_\varphi} = \frac{\boldsymbol{e_m} \times \boldsymbol{m_{1,\parallel xy}}}{|\boldsymbol{e_m} \times \boldsymbol{m_{1,\parallel xy}}|}, \qquad \boldsymbol{e_\theta} = \frac{\boldsymbol{e_\varphi} \times \boldsymbol{e_m}}{|\boldsymbol{e_\varphi} \times \boldsymbol{e_m}|}, \qquad \text{(S8)}$$

where $\boldsymbol{m_{1,\parallel xy}}$ is the in-plane component of $\boldsymbol{m_1^{eq}}$. Since the phase between $m_1$ and $m_2$ is constant when magnetization precession is stable, we consider the joint precession modes (AM and OM) of $m_1$ and $m_2$. First, we define a new $\boldsymbol{z}$'-axis (the bisector of two magnetization vectors): $\boldsymbol{z}' \parallel \boldsymbol{m_1^{eq}} + \boldsymbol{m_2^{eq}}$. An operator $C_{2z'}$ is introduced to represent a 180° rotation about the $\boldsymbol{z}$'-axis, yielding $\boldsymbol{m_2^{eq}} = C_{2z'}\boldsymbol{m_1^{eq}}$. Therefore, the acoustic mode can be represented as $\delta\boldsymbol{m_+} = \delta\boldsymbol{m_1} + C_{2z'}\delta\boldsymbol{m_2}$, corresponding to the in-phase motion of two magnetic moments, while the optical mode is $\delta\boldsymbol{m_-} = \delta\boldsymbol{m_1} - C_{2z'}\delta\boldsymbol{m_2}$, corresponding to the counter-phase motion of magnetic moments, as shown in Fig. S2.

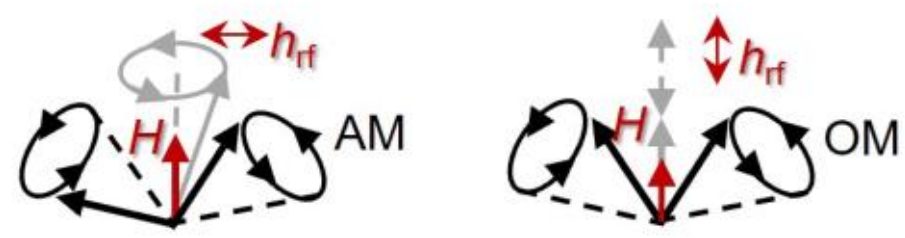

Figure S2. Schematic illustrations of magnetization in-phase precession for the in-phase acoustic mode (AM) and out-of-phase optical mode (OM).

Using these transformations, we obtain the eigenvalue equation in the local coordinate system

$$-\frac{i\omega}{\gamma}\begin{pmatrix}\delta\boldsymbol{m}_{+\boldsymbol{\varphi}}\\ \delta\boldsymbol{m}_{+\boldsymbol{\theta}}\\ \delta\boldsymbol{m}_{+\mathbf{m}}\\ \delta\boldsymbol{m}_{-\boldsymbol{\varphi}}\\ \delta\boldsymbol{m}_{-\boldsymbol{\theta}}\\ \delta\boldsymbol{m}_{-\mathbf{m}}\end{pmatrix} = D\cdot A\cdot D^{-1}\begin{pmatrix}\delta\boldsymbol{m}_{+\boldsymbol{\varphi}}\\ \delta\boldsymbol{m}_{+\boldsymbol{\theta}}\\ \delta\boldsymbol{m}_{+\mathbf{m}}\\ \delta\boldsymbol{m}_{-\boldsymbol{\varphi}}\\ \delta\boldsymbol{m}_{-\boldsymbol{\theta}}\\ \delta\boldsymbol{m}_{-\mathbf{m}}\end{pmatrix}, \qquad D=\begin{pmatrix}P & C_{2z'}P\\ P & -C_{2z'}P\end{pmatrix}. \quad \text{(S9)}$$

The off-diagonal block elements of the matrix $D\cdot A\cdot D^{-1}$ represent the coupling terms between the AM and OM. By setting the off-diagonal block matrix elements to zero, we can obtain the pure acoustic and optical modes.

We now consider the low-field right-handed (RM) and left-handed (LM) coupling regimes before the spin-flop transition. Distinguishing between right- and left-handed precession requires a global reference (external magnetic field *H*) to define the rotation direction. The local basis vectors are represented as

$$\boldsymbol{e_m}=\frac{\boldsymbol{m_1^{eq}}}{\left|\boldsymbol{m_1^{eq}}\right|}, \qquad \boldsymbol{e_\varphi}=\frac{\boldsymbol{e_m}\times\boldsymbol{H}}{|\boldsymbol{e_m}\times\boldsymbol{H}|}, \qquad \boldsymbol{e_\theta}=\frac{\boldsymbol{e_\varphi}\times\boldsymbol{e_m}}{|\boldsymbol{e_\varphi}\times\boldsymbol{e_m}|}, \qquad \text{(S10)}$$

The right-handed mode (RM) is defined as $\delta\boldsymbol{m_{j+}}=\delta\boldsymbol{m_{j\varphi}}+i\delta\boldsymbol{m_{j\theta}}, j=1,2$, corresponding to the right circular precession of both magnetic moments; while the left-handed mode (LM) is defined as $\delta\boldsymbol{m_{j-}}=\delta\boldsymbol{m_{j\varphi}}-i\delta\boldsymbol{m_{j\theta}}, j=1,2$, corresponding to the left circular precession, as shown in Fig. S3, where $\delta\boldsymbol{m}_{j\varphi}$ and $\delta\boldsymbol{m}_{j\theta}$ represent the orthogonal in-plane components of the magnetic precession.

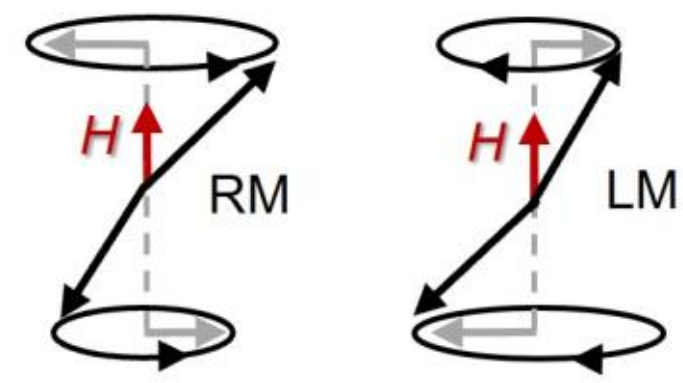

Figure S3. Schematic illustrations of magnetization precession for right-handed (RM) and left-handed modes (LM).

The rotation matrix $C_{2z'}$ is again employed to ensure the parallel configuration of the two sublattices. Using these definitions and transformations, we derive the eigenvalue equation for the resonance frequencies in this local system.

$$-\frac{i\omega}{\gamma}\begin{pmatrix}\delta\boldsymbol{m}_{1+}\\ \delta\boldsymbol{m}_{2+}\\ \delta\boldsymbol{m}_{1m}\\ \delta\boldsymbol{m}_{1-}\\ \delta\boldsymbol{m}_{2-}\\ \delta\boldsymbol{m}_{2m}\end{pmatrix} = E\cdot A\cdot E^{-1}\begin{pmatrix}\delta\boldsymbol{m}_{1+}\\ \delta\boldsymbol{m}_{2+}\\ \delta\boldsymbol{m}_{1m}\\ \delta\boldsymbol{m}_{1-}\\ \delta\boldsymbol{m}_{2-}\\ \delta\boldsymbol{m}_{2m}\end{pmatrix}, E = \begin{pmatrix}1 & i & 0 & 0 & 0 & 0\\ 0 & 0 & 0 & 1 & i & 0\\ 0 & 0 & 1 & 0 & 0 & 0\\ 1 & -i & 0 & 0 & 0 & 0\\ 0 & 0 & 0 & 1 & -i & 0\\ 0 & 0 & 0 & 0 & 0 & 1\end{pmatrix}\begin{pmatrix}P & 0\\ 0 & C_{2z'}P\end{pmatrix}. \quad (S11)$$

The off-diagonal block elements of the matrix $E\cdot A\cdot E^{-1}$ represent the coupling terms between the RM and LM. By setting the off-diagonal block matrix elements to zero, we can obtain the pure left-handed and right-handed modes.

## Section III. Theoretical derivation of the quantum model

### A. AM-OM coupling case

For a ferromagnetic spin configuration, the magnetization operators can be quantized using the Holstein-Primakoff (HP) transformation [8–10], introducing the creation and annihilation operators for the two sublattices, $a_1$ and $a_2$:

$$\widehat{M}_{1\varphi} - i\widehat{M}_{1\theta} = \sqrt{2\gamma_0\hbar M_s}\,\hat{a}_1$$

$$\widehat{M}_{2\varphi} - i\widehat{M}_{2\theta} = \sqrt{2\gamma_0\hbar M_s}\,\hat{a}_2$$

$$\hat{M}_{1\varphi} + i\hat{M}_{1\theta} = \sqrt{2\gamma_0 \hbar M_s}\hat{a}_1^\dagger$$

$$\hat{M}_{2\varphi} + i\hat{M}_{2\theta} = \sqrt{2\gamma_0 \hbar M_s}\hat{a}_2^\dagger$$

$$\hat{M}_{1m} = M_s - \gamma_0 \hbar \hat{a}_1^\dagger \hat{a}_1$$

$$\hat{M}_{2m} = M_s - \gamma_0 \hbar \hat{a}_2^\dagger \hat{a}_2 \,. \quad \text{(S12)}$$

Consistent with the definitions in the classical LL model, we introduce mode operators formed from linear combinations of sublattice magnetizations. In the high-field regime, we define $\hat{M}_\pm = \hat{M}_1 \pm C_{2z'}\hat{M}_2$ to describe the AM-OM coupling case. The Hamiltonian can then be obtained by Eq. (S2). Since the magnetization three-component operators are dependent, we can eliminate the *z*-component of magnetization using the normalization condition, and consider only quadratic terms. This allows the exact separation of the Hamiltonian into pure $\hat{M}_+$ or $\hat{M}_-$ mode magnetization terms and coupled mode magnetization terms with products of $\hat{M}_+$ and $\hat{M}_-$ components. For the AM-OM case, the Hamiltonian can be decomposed into pure mode and coupled mode parts:

$$\hat{\mathcal{H}}_{all} = \hat{\mathcal{H}}_{pure} + \hat{\mathcal{H}}_{couple} \,, \qquad \text{(S13)}$$

$$\hat{\mathcal{H}}_{pure} = A\left(\hat{a}_1^\dagger \hat{a}_1 + \hat{a}_2^\dagger \hat{a}_2\right) + B(\hat{a}_1^2 + \hat{a}_2^2) + C\hat{a}_1\hat{a}_2 + D\hat{a}_1^\dagger \hat{a}_2 + h.c., \text{(S14)}$$

$$\hat{\mathcal{H}}_{couple} = E\hat{a}_1^\dagger \hat{a}_1 + F\hat{a}_2^\dagger \hat{a}_2 + G\hat{a}_1^2 + H\hat{a}_2^2 + I\hat{a}_1\hat{a}_2 + J\hat{a}_1^\dagger \hat{a}_2 + h.c. \,, \text{(S15)}$$

where all the coefficients are real-valued.

We then perform the Bogoliubov transformation to diagonalize $\hat{\mathcal{H}}_{pure}$. The annihilation (creation) operators $\hat{a}_+$ ($\hat{a}_+^\dagger$) and $\hat{a}_-$ ($\hat{a}_-^\dagger$) correspond to the acoustic and optical magnon modes in the AM-OM case:

$$\hat{a}_\pm = w_\pm \hat{a}_1 + x_\pm \hat{a}_2 + y_\pm \hat{a}_1^\dagger + z_\pm \hat{a}_2^\dagger \,. \quad \text{(S16)}$$

The boson commutation rules require the coefficients to satisfy the normalization condition

$$w_i w_j^* + x_i x_j^* - y_i y_j^* - z_i z_j^* = \delta_{ij}, i(j) \in \{+,-\}\,. \quad (S17)$$

Through the inverse transformation of Eq. (S16), we can express $\hat{\mathcal{H}}_{pure}$ as represented, like Eq. (S14), in terms of $\hat{a}_{\pm}$. The Heisenberg equations of motion are

$$\frac{d\hat{a}_{\pm}}{dt} = -i\omega_{\pm}\hat{a}_{\pm} = \frac{1}{i\hbar}\left[\hat{a}_{\pm}, \hat{\mathcal{H}}_{pure}\right]. \quad (S18)$$

Through the HP transformation, Eq. (S18) can be reformulated as a fourth-order matrix equation, and the eigenfrequencies and eigenvectors can be obtained by solving this equation

$$\hbar\omega_{\pm}\begin{bmatrix} w_{\pm} \\ x_{\pm} \\ y_{\pm} \\ z_{\pm} \end{bmatrix} = \begin{bmatrix} 2A & D & -2B & -C \\ D & 2A & -C & -2B \\ 2B & C & -2A & -D \\ C & 2B & -D & -2A \end{bmatrix}\begin{bmatrix} w_{\pm} \\ x_{\pm} \\ y_{\pm} \\ z_{\pm} \end{bmatrix}. \quad (S19)$$

The obtained eigenvectors (all nonzero) are used to express the coefficients in the Bogoliubov transformation [Eq. (S16)], we can similarly represent $\hat{\mathcal{H}}_{couple}$ of Eq. (S15) in terms of $\hat{a}_{\pm}$, and then reorganize the coefficients. By employing the bosonic commutation relations, the diagonalized $\hat{\mathcal{H}}_{pure}$ can be written as the sum of two decoupled + (AM) and – (OM) harmonic oscillators, while the $\hat{\mathcal{H}}_{couple}$ can be formulated as cross terms between the + (AM) and − (OM) operators

$$\hat{\mathcal{H}}_{pure} = \hbar\omega_+\left(\hat{a}_+^{\dagger}\hat{a}_+ + \frac{1}{2}\right) + \hbar\omega_-\left(\hat{a}_-^{\dagger}\hat{a}_- + \frac{1}{2}\right),$$

$$\hat{\mathcal{H}}_{couple} = i\hbar g_1\left(\hat{a}_+\hat{a}_-^{\dagger} - \hat{a}_+^{\dagger}\hat{a}_-\right) + i\hbar g_2\left(\hat{a}_+^{\dagger}\hat{a}_-^{\dagger} - \hat{a}_+\hat{a}_-\right). \quad (S20)$$

For $\hat{\mathcal{H}}_{pure}$, $\omega_+$ and $\omega_-$ correspond to the frequencies of pure AM and OM, respectively. For $\hat{\mathcal{H}}_{couple}$, the co-rotating coupling term $g_1$ represents the magnon exchange leading to AM-OM hybridization, while the counter-rotating coupling $g_2$ arises from the creation or annihilation of AM-OM magnon pairs, resulting in a two-mode squeezing state. To figure out the influence of $g_1$ and $g_2$, and obtain the frequency dispersions of the coupled modes, we need to introduce coupled magnon annihilation operators: $\hat{a}_{\mathrm{H}}$ ($\hat{a}_{\mathrm{L}}$) corresponds to the annihilation operator of the high-frequency mode (HFM) (low-frequency mode (LFM)), and $\hat{a}_{\mathrm{H}}^{\dagger}$ ($\hat{a}_{\mathrm{L}}^{\dagger}$) corresponds to the creation operator of the HFM (LFM) magnon mode. The total Hamiltonian Eq. (S20) can be diagonalized by the Hopfield-Bogoliubov transformation:

$$\widehat{\mathcal{H}}_{all} = \hbar\omega_{\mathrm{H}}\left(\hat{a}_{\mathrm{H}}^{\dagger}\hat{a}_{\mathrm{H}} + \frac{1}{2}\right) + \hbar\omega_{\mathrm{L}}\left(\hat{a}_{\mathrm{L}}^{\dagger}\hat{a}_{\mathrm{L}} + \frac{1}{2}\right),$$

$$\hat{a}_i = w_i\hat{a}_+ + x_i\hat{a}_- + y_i\hat{a}_+^{\dagger} + z_i\hat{a}_-^{\dagger}\,, i = \mathrm{H},\mathrm{L}\,, \quad (S21)$$

where $\omega_{\mathrm{H}}$ ($\omega_{\mathrm{L}}$) corresponds to the eigenfrequency of HFM (LFM). The normalization condition constrains the coefficients

$$w_{\mathrm{i}}w_j^* + x_{\mathrm{i}}x_j^* - y_{\mathrm{i}}y_j^* - z_{\mathrm{i}}z_j^* = \delta_{ij}, i(j) \in \{\mathrm{H},\mathrm{L}\}\,. \quad (S22)$$

We solve the Heisenberg equations of motion, the same as Eq. (S18), and get the matrix equation

$$\omega_{\mathrm{H(L)}}\begin{bmatrix} w_{\mathrm{H(L)}} \\ x_{\mathrm{H(L)}} \\ y_{\mathrm{H(L)}} \\ z_{\mathrm{H(L)}} \end{bmatrix} = \begin{bmatrix} \omega_+ & ig_1 & 0 & ig_2 \\ -ig_1 & \omega_- & ig_2 & 0 \\ 0 & ig_2 & -\omega_+ & ig_1 \\ ig_2 & 0 & -ig_1 & -\omega_- \end{bmatrix}\begin{bmatrix} w_{\mathrm{H(L)}} \\ x_{\mathrm{H(L)}} \\ y_{\mathrm{H(L)}} \\ z_{\mathrm{H(L)}} \end{bmatrix}. \quad (S23)$$

The obtained eigenfrequency $\omega_{\mathrm{H}}$ ($\omega_{\mathrm{L}}$) corresponds to the $\omega$ obtained from the classical LL model in Section II. Using Eq. (S23), we can evaluate and discuss mode squeezing, i.e., the vacuum Rabi splitting-induced shift (VRSS) and vacuum Bloch–Siegert shift (VBSS), by selecting $g_1$ or $g_2$ to zero.

### B. RM-LM coupling case

For an antiferromagnetic two-sublattice configuration, the ladder operators for the $a_2$ sublattice are reversed in the HP transformation, as follows:

$$\widehat{M}_{1\varphi} - i\widehat{M}_{1\theta} = \sqrt{2\gamma_0\hbar M_s}\hat{a}_1$$

$$\widehat{M}_{2\varphi} - i\widehat{M}_{2\theta} = \sqrt{2\gamma_0\hbar M_s}\hat{a}_2^{\dagger}$$

$$\widehat{M}_{1\varphi} + i\widehat{M}_{1\theta} = \sqrt{2\gamma_0\hbar M_s}\hat{a}_1^{\dagger}$$

$$\widehat{M}_{2\varphi} + i\widehat{M}_{2\theta} = \sqrt{2\gamma_0\hbar M_s}\hat{a}_2$$

$$\widehat{M}_{1m} = M_s - \gamma_0\hbar\hat{a}_1^\dagger\hat{a}_1$$

$$\widehat{M}_{2m} = -M_s + \gamma_0\hbar\hat{a}_2^\dagger\hat{a}_2\,. \quad \text{(S24)}$$

For the low-field AFM before the spin-flop transition, we define local chiral combinations $\widehat{M}_\pm = \widehat{M}_{j\varphi} \pm i\widehat{M}_{j\theta}, j = 1{,}2,$ as RM and LM. Since the component $\widehat{M}_m$ operators incorporate crucial equilibrium-position information; the separation of pure and coupled Hamiltonians is performed after applying the HP transformation. For the RM-LM coupling case, the Hamiltonian can also be decomposed into

$$\widehat{\mathcal{H}}_{pure} = A_1\hat{a}_1^\dagger\hat{a}_1 + A_2\hat{a}_2^\dagger\hat{a}_2 + A_3\hat{a}_1\hat{a}_1^\dagger + A_4\hat{a}_2\hat{a}_2^\dagger + C_1\hat{a}_1\hat{a}_2 + C_3\hat{a}_1^\dagger\hat{a}_2^\dagger\,, \text{(S25)}$$

$$\widehat{\mathcal{H}}_{couple} = B_1\hat{a}_1^2 + B_2\hat{a}_2^2 + B_3a_1^{\dagger 2} + B_4a_2^{\dagger 2} + C_1\hat{a}_1\hat{a}_2 + C_3\hat{a}_1^\dagger\hat{a}_2^\dagger\,. \text{(S26)}$$

We then apply the Bogoliubov transformation given in Eq. (S16) to diagonalize the $\widehat{\mathcal{H}}_{pure}$. Here, the annihilation (creation) operators $\hat{a}_+$ $(\hat{a}_+^\dagger)$ and $\hat{a}_-$ $(\hat{a}_-^\dagger)$ correspond to the right-handed (RM) and left-handed (LM) magnon modes for the RM-LM coupling case. We can also get the matrix equation similar to Eq. (S19)

$$\hbar\omega_\pm\begin{bmatrix} w_\pm \\ x_\pm \\ y_\pm \\ z_\pm \end{bmatrix} = \begin{bmatrix} A_1 + A_3 & D_3 & -2B_1 & -C_1 \\ D_1 & A_2 + A_4 & -C_1 & -2B_2 \\ 2B_3 & C_3 & -A_1 - A_3 & -D_1 \\ C_3 & 2B_4 & -D_3 & -A_2 - A_4 \end{bmatrix}\begin{bmatrix} w_\pm \\ x_\pm \\ y_\pm \\ z_\pm \end{bmatrix}. \quad \text{(S27)}$$

For the RM-LM case, the coefficients in the Bogoliubov transformation satisfy $w_+ = x_-$, $z_+ = y_-$ and all other coefficients being zero. Through numerical analysis and selecting an appropriate phase, we can simplify the Hamiltonian to the following form

$$\widehat{\mathcal{H}}_{pure} = \hbar\omega_+\hat{a}_+^\dagger\hat{a}_+ + \hbar\omega_-\hat{a}_-^\dagger\hat{a}_-\,,$$

$$\widehat{\mathcal{H}}_{couple} = i\hbar g_1\left(\hat{a}_+\hat{a}_-^\dagger - \hat{a}_+^\dagger\hat{a}_-\right) + \hbar\left[g_{3+}\left(\hat{a}_+^2 + \hat{a}_+^{\dagger\,2}\right) - g_{3-}\left(\hat{a}_-^2 + \hat{a}_-^{\dagger\,2}\right)\right]. \quad \text{(S28)}$$

For $\widehat{\mathcal{H}}_{pure}$, $\omega_+$ and $\omega_-$ correspond to the frequencies of pure RM and LM, respectively. For $\widehat{\mathcal{H}}_{couple}$, besides the co-rotating coupling $g_1$, the self-squared couplings $g_{3+}$, $g_{3-}$ due to the creation and annihilation of RM-RM and LM-LM magnon pairs, leading to a single-mode squeezed state. It differs from the counter-rotating coupling $g_2$-induced two-mode squeezing state in the AM-OM coupling case. To obtain the mode squeezing controlled by coupling terms $g_1$, $g_{3+}$, and $g_{3-}$, and the frequencies of the coupled modes, we again introduce the coupled magnon operators as defined in Eq. (S21). This yields a matrix equation similar to Eq. (S23)

$$\omega_{\mathrm{H(L)}}\begin{bmatrix} w_{\mathrm{H(L)}} \\ x_{\mathrm{H(L)}} \\ y_{\mathrm{H(L)}} \\ z_{\mathrm{H(L)}} \end{bmatrix} = \begin{bmatrix} \omega_+ & ig_1 & -2g_{3+} & 0 \\ -ig_1 & \omega_- & 0 & -2g_{3-} \\ 2g_{3+} & 0 & -\omega_+ & ig_1 \\ 0 & 2g_{3-} & -ig_1 & -\omega_- \end{bmatrix}\begin{bmatrix} w_{\mathrm{H(L)}} \\ x_{\mathrm{H(L)}} \\ y_{\mathrm{H(L)}} \\ z_{\mathrm{H(L)}} \end{bmatrix}. \quad \text{(S29)}$$

The results of VRSS and VBSS are shown and discussed in the main text.

### C. Quantification of quadrature fluctuations and magnon squeezing states

The theoretical framework presented in the previous section establishes the foundation for understanding and quantifying the quantum squeezing inherent in the ultrastrongly coupled ground state. As described in the main text, the entanglement of the vacuum state due to counter-rotating ($g_2$) and self-squared ($g_3$) interactions leads to a redistribution of quantum noise between conjugate variables, i.e., squeezing. To quantify this, we define a general quadrature operator constructed from the original sublattice magnon operators

To confirm and quantify the squeezing in these hybrid magnon modes, we evaluate the quantum fluctuations of the ground state in the selected quadratures. First, we define a general quadrature operator constructed from sublattice ladder operators $\hat{a}_{1,2}$ (defined in Eqs. (S12) and (S24)), parameterized by two independent phases $\Phi$ and $\psi$: [10,11]

$$\hat{X}_{\phi,\psi} = \frac{1}{\sqrt{8}}\left[\hat{a}_1 e^{-i\phi} + \hat{a}_1^\dagger e^{i\phi} + \hat{a}_2 e^{-i\psi} + \hat{a}_2^\dagger e^{i\psi}\right]. \quad \text{(S30)}$$

Its variance (quantum fluctuation) in an arbitrary quantum state is defined as

$$\langle \delta\hat{X}_{\phi,\psi}^2 \rangle = \langle \hat{X}_{\phi,\psi}^2 \rangle - \langle \hat{X}_{\phi,\psi} \rangle^2. \quad \text{(S31)}$$

Using the bosonic commutation relations, one finds $\left[\hat{X}_{\phi,\psi}, \hat{X}_{\phi+\frac{\pi}{2},\psi+\frac{\pi}{2}}\right] = \frac{i}{2}$. The Heisenberg uncertainty principle therefore imposes the constraint

$$\langle \delta\hat{X}_{\phi,\psi}^2 \rangle \cdot \left\langle \delta\hat{X}_{\phi+\frac{\pi}{2},\psi+\frac{\pi}{2}}^2 \right\rangle \geq \left( \frac{1}{2i} \left\langle \left[\hat{X}_{\phi,\psi}, \hat{X}_{\phi+\frac{\pi}{2},\psi+\frac{\pi}{2}}\right] \right\rangle \right)^2 = \frac{1}{16}. \quad \text{(S32)}$$

For the uncoupled, unsqueezed coherent state, such as the ground state $|0_1, 0_2\rangle$ of the pure Hamiltonian $\hat{\mathcal{H}}_{pure}$ in Eqs. (S14) and (S25), the variance for any quadrature exhibits the minimum theoretically allowed value: $\langle 0_1, 0_2 | \delta\hat{X}_{\phi,\psi}^2 | 0_1, 0_2 \rangle = \frac{1}{4}$, which corresponds to a standard quantum limit (SQL) of $\frac{1}{4}$ for an unsqueezed coherent state. In contrast, for the entangled ground state $|0_{\mathrm{H}}, 0_{\mathrm{L}}\rangle$ of the full Hamiltonian $\hat{\mathcal{H}}_{all}$ in Eq. (S21), the quadrature fluctuation is calculated by expressing the sublattice operators in terms of the final hybrid operators. This involves applying the inverse of the Bogoliubov transformations: first, use the inverse of Eq. (S16) to express $\hat{a}_{1,2}$ via $\hat{a}_{+,-}$; then, use the inverse of Eq. (S21) to express $\hat{a}_{+,-}$ via $\hat{a}_{\mathrm{H,L}}$. Substituting these relationships into Eqs. (S30) and (S31) yields the fluctuation in the coupled ground state $|0_{\mathrm{H}}, 0_{\mathrm{L}}\rangle$ as

$$\langle 0_{\mathrm{H}}, 0_{\mathrm{L}} | \delta\hat{X}_{\phi,\psi}^2 | 0_{\mathrm{H}}, 0_{\mathrm{L}} \rangle = \frac{1}{8}(A_H A_H^* + A_L A_L^*). \quad \text{(S33)}$$

where the complex coefficients are given by

$$A_H = A_+ w_{\mathrm{H}}^* + A_- x_{\mathrm{H}}^* - A_+^* y_{\mathrm{H}}^* - A_-^* z_{\mathrm{H}}^*,$$

$$A_L = A_+ w_{\mathrm{L}}^* + A_- x_{\mathrm{L}}^* - A_+^* y_{\mathrm{L}}^* - A_-^* z_{\mathrm{L}}^*,$$

$$A_+ = w_+^* e^{-i\phi} + x_+^* e^{-i\psi} - y_+^* e^{i\phi} - z_+^* e^{i\psi},$$

$$A_- = w_-^* e^{-i\phi} + x_-^* e^{-i\psi} - y_-^* e^{i\phi} - z_-^* e^{i\psi}.$$

Here, $w_\pm$, $x_\pm$, $y_\pm$, $z_\pm$ and $w_{\mathrm{H,L}}$, $x_{\mathrm{H,L}}$, $y_{\mathrm{H,L}}$, $z_{\mathrm{H,L}}$ are the coefficients from the first and second Bogoliubov transformations (see explicit forms in previous text), respectively.

By varying the phases $\Phi$ and $\psi$ from 0 to 360°, we obtain the quadrature fluctuation landscape. Two specific conjugate quadratures corresponding to the principal squeezing axes [11] are found for $\Phi = \psi = 0$ and $\Phi = \psi = 90°$ as

$$\hat{X}_1 = \frac{1}{\sqrt{8}}\left[\hat{a}_1 + \hat{a}_1^\dagger + \hat{a}_2 + \hat{a}_2^\dagger\right],$$

$$\hat{X}_2 = \frac{1}{i\sqrt{8}}\left[\hat{a}_1 - \hat{a}_1^\dagger + \hat{a}_2 - \hat{a}_2^\dagger\right]. \quad \text{(S34)}$$

For the squeezed ground state $|0_{\mathrm{H}}, 0_{\mathrm{L}}\rangle$, $\langle \delta\hat{X}_1^2 \rangle$ is minimized, while $\langle \delta\hat{X}_2^2 \rangle$ is maximized, still satisfying the uncertainty relation $\langle \delta\hat{X}_1^2 \rangle \cdot \langle \delta\hat{X}_2^2 \rangle \geq \frac{1}{16}$, but individually deviating from the SQL of $\frac{1}{4}$. A value of $\langle \delta\hat{X}_1^2 \rangle < \frac{1}{4}$ directly indicates quadrature squeezing, with smaller values corresponding to stronger squeezing.

A key achievement of our work is linking these abstract quadrature operators to physically measurable quantities, thereby establishing a protocol for experimental detection. Using the HP transformations (Eq. (S12) for AM-OM and Eq. (S24) for RM-LM), the quadratures $\hat{X}_1$ and $\hat{X}_2$ can be mapped onto linear combinations of the transverse magnetization components of the two sublattices. Within a right-handed coordinate system defined relative to the equilibrium magnetization plane (see Fig. S4), this mapping takes a unified form for both AM-OM and RM-LM coupling regimes as

$$\hat{X}_1 = \frac{1}{\sqrt{2}\sqrt{2\gamma_0 \hbar M_s}}\left[\hat{M}_{1\varphi} + \hat{M}_{2\varphi}\right],$$

$$\hat{X}_2 = \frac{-1}{\sqrt{2}\sqrt{2\gamma_0 \hbar M_s}}\left[\hat{M}_{1\theta} + \hat{M}_{2\theta}\right]. \quad \text{(S35)}$$

Here, $\widehat{M}_{i\theta}$ and $\widehat{M}_{i\varphi}$ are the operators for the transverse magnetization components for sublattice $i$ ($i$ = 1, 2) along the out-of-plane and in-plane directions relative to the equilibrium $M_1$-$M_2$ plane, respectively. $\gamma_0$ is the gyromagnetic ratio, and $M_s$ is the saturation magnetization.

The schematic in Figs. S4(a) and (b) clarifies the physical meaning of these mapped observables in two magnetic configurations. For the RM-LM configuration (low field, AFM$_c$ phase), $\boldsymbol{X_1}$ relates to the in-plane component of the total spin, while $\boldsymbol{X_2}$ relates to the out-of-plane component of the Néel order. For the AM-OM configuration (high field, canted phase), both $\boldsymbol{X_1}$ and $\boldsymbol{X_2}$ correspond to the in-plane and out-of-plane components of the Néel vector, respectively (see Fig. S4 caption for details).

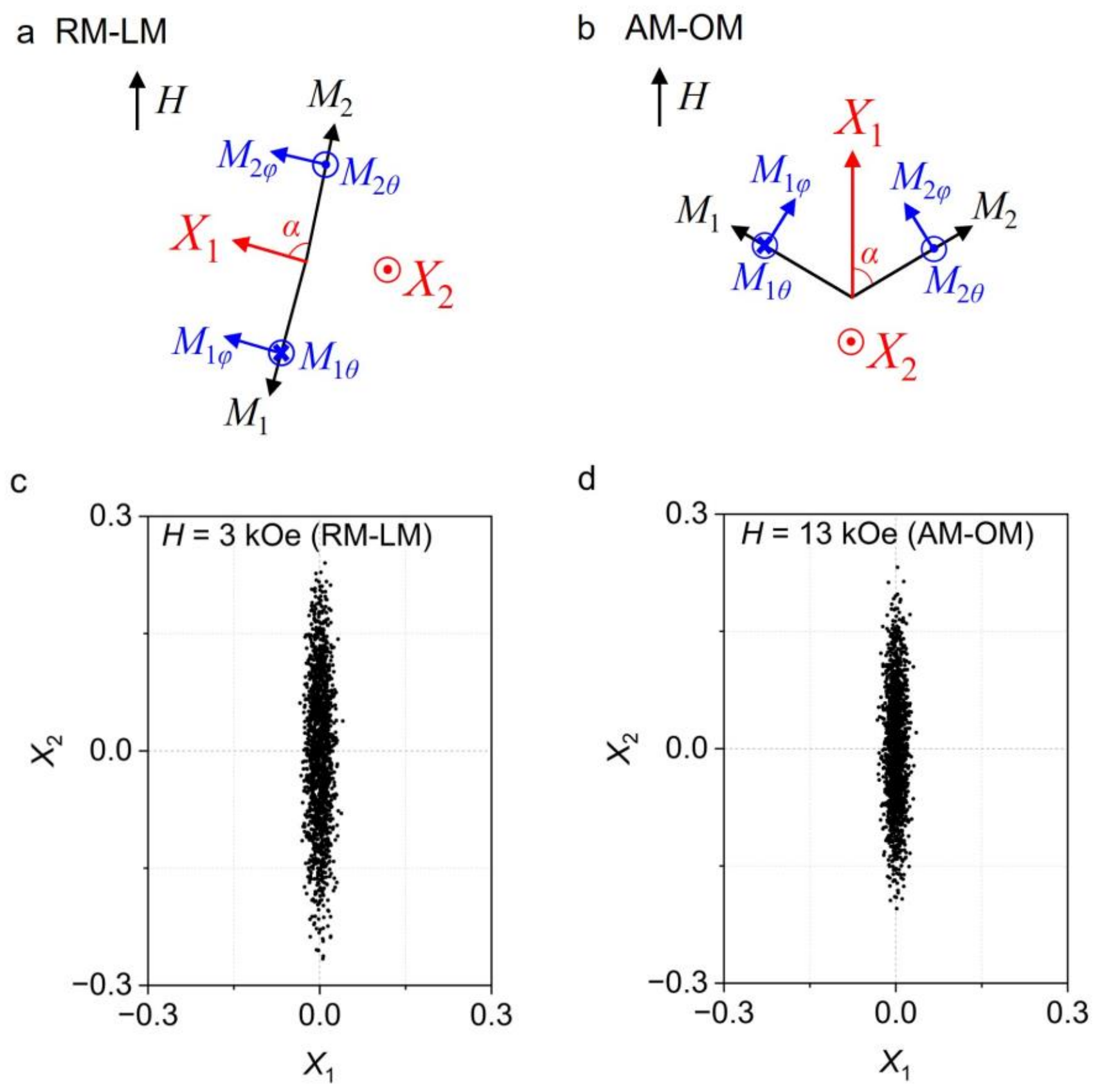

**Figure S4.** Schematic definitions of the coordinate systems and representative values of the quadratures $\boldsymbol{X_1}$ and $\boldsymbol{X_2}$ for the RM-LM and AM-OM configurations. (a, b) Definitions of the local coordinate systems for the two sublattices in the RM-LM and AM-OM configurations, respectively. $M_{1\theta}$ and $M_{2\theta}$ are antiparallel and perpendicular to the $M_1$-$M_2$ plane. Both $M_{1\varphi}$ and $M_{2\varphi}$ form right-

handed coordinate systems. Here $\alpha$ is half of the angle between the equilibrium magnetization directions $M_1$ and $M_2$ of the two sublattices. (c) Representative distribution statistics of $\boldsymbol{X_1}$ and $\boldsymbol{X_2}$ values at $H$ = 3 kOe in the RM-LM regime. (d) Representative distribution statistics of $\boldsymbol{X_1}$ and $\boldsymbol{X_2}$ values at $H$ = 13 kOe in the AM-OM regime.

While the theoretical mapping is exact, measuring the quantum variance $\langle \delta\hat{X}_1^2 \rangle$, $\langle \delta\hat{X}_2^2 \rangle$ directly in an experiment is challenging. Here, we demonstrate a powerful, general protocol that bridges this gap: we use stochastic, thermally-driven micromagnetic simulations (Section IV) to extract these quantum-level fluctuations from the simulated, classical magnetization dynamics. The connection is established via the correspondence principle and the quantum-to-classical transition for linear spin-wave operators. The variance of the quantum operator $\hat{X}_1$ is proportional to the expectation value of $\langle \delta\hat{X}_1^2 \rangle$ in the ground state. In the semi-classical limit relevant to our simulations, this corresponds to the time-averaged mean-square fluctuation of the corresponding classical observable $X_1(t)$, constructed from the simulated sublattice magnetizations $M_i(t)$ ($i$ = 1, 2) via the mapping in Eq. (S35):

$$\langle \delta\hat{X}_1^2 \rangle_{sim} \equiv \overline{(X_1(t) - \overline{X_1})^2} \propto \langle 0_\mathrm{H}, 0_\mathrm{L} | \delta\hat{X}_{\phi,\psi}^2 | 0_\mathrm{H}, 0_\mathrm{L} \rangle . \quad (S36)$$

Here, the overline denotes a time average over the simulated, thermally fluctuating magnetization trajectories. Our simulations are performed at a finite but very low temperature (0.1 K) to seed the fluctuations while remaining in the linear, small-amplitude spin-wave regime, effectively probing the ground-state quantum fluctuations. As illustrated in Figs. S4(c, d), the representative distributions of $X_1$ and $X_2$ values sampled from the micromagnetic simulation trajectories are highly anisotropic. The variance along the squeezed quadrature $X_1$ is significantly smaller than that along the anti-squeezed quadrature $X_2$, providing a direct visual confirmation of the noise redistribution predicted by the quantum model.

Finally, the degree of noise suppression is quantified by the squeezing factor $S_{\mathrm{sq}}$, defined in decibels (dB) as

$$S_{\mathrm{sq}} = -10\log_{10}\left(\frac{\langle \delta \hat{X}_1^2 \rangle}{0.25}\right). \quad (S37)$$

A positive value ($S_{\mathrm{sq}} > 0$ dB) indicates squeezing below the SQL, with larger values signifying stronger suppression. This framework provides a complete theoretical protocol for detecting and quantifying magnon squeezing, which is validated against the micromagnetic simulation results in the main text [Fig. 3].

**Section IV. Micromagnetic simulation results**

The micromagnetic simulations were performed using Mumax[3] by solving the Landau–Lifshitz–Gilbert (LLG) equation [12,13]. The simulation dimensions of $CrPS_4$ were set to 10 × 10 × 2 grid with a cell size of 1.1 × 0.7 × 0.6 nm$^3$ and periodic boundary conditions of 100 × 200 × 50 [1,2]. The simulation parameters, including the exchange stiffness constant $A_{\mathrm{ex}}$ = 1.30 × 10$^{-12}$ J/m, saturation magnetization $M_{\mathrm{sat}}$ = 7.28 × 10$^4$ A/m, the RKKY coupling coefficient $J = \frac{\mu_0 M_s H_E t}{2}$ = -0.66 × 10$^{-4}$ J/m$^2$, the uniaxial anisotropy in the *c*-axis $K_{\mathrm{m}} = -\frac{\mu_0 M_s H_{Am}}{2sin\beta cos\beta}$ = 5.27 × 10$^4$ J/m$^3$, in the *a*-axis $K_1 = -\frac{\mu_0 M_s H_{A1}}{2} - K_m(cos^2\beta - sin^2\beta)$ = 4.48 × 10$^4$ J/m$^3$, and in the *b*-axis $K_2 = -\frac{\mu_0 M_s H_{A2}}{2} + K_m sin^2\beta$ = 4.89 × 10$^4$ J/m$^3$, are determined based on the fitting parameters of the AFMR experiment at 7 K [Fig. 1(h)]. Firstly, the static magnetization is achieved by minimizing the system's energy under a given bias field $H$. Secondly, starting from the obtained static magnetization, the time evolution of the spatially averaged magnetization $\boldsymbol{m}(t)$ is recorded with a perturbation field $h_{\mathrm{rf}}(t) = h_0\sin(2\pi f_c t)/(2\pi f_c t)$ along both the *x*- and *y*-axes to excite all dynamic modes, with an amplitude $h_0$ of 5 mT and a cut-off frequency $f_c$ of 100 GHz. The magnon spectra can be derived from the Fourier transform of $\boldsymbol{m}(t)$ with $H$ ranging from 0 to 14 kOe in steps of 0.2 kOe.

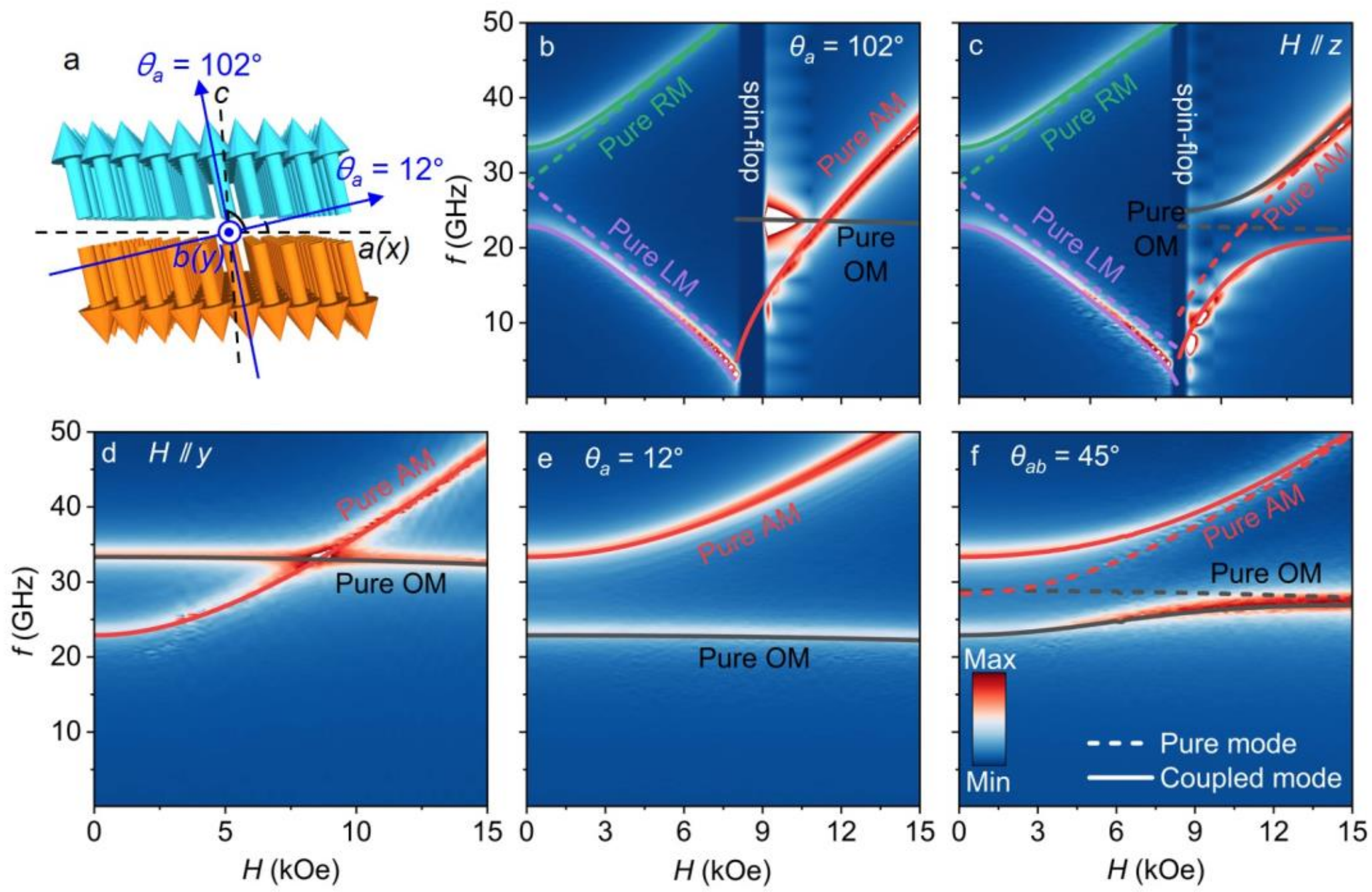

**Figure S5.** Simulated and theoretical dispersions of the dynamic modes with the parameters obtained by our experiments at 7 K. (a) Schematic illustration of the spin configuration of the simulated $CrPS_4$ with three crystal axes (*a*-, *b*- and *c*-axes) in the Cartesian coordinate system (*x*, *y*, *z*). (b-f) The AFMR *f*-*H* spectra obtained by micromagnetic simulations (color mapping) and microscopic LL model (lines) under different *H* angles: (b) $\theta_a$ = 102°, (c) *H* ∥ *z*, (d) *H* ∥ *y*, (e) $\theta_a$ = 12°, and in-plane angle between the *a* and *b* axes (f) $\theta_{ab}$ = 45°. The solid lines represent the complete spectral modes, and the dotted lines represent the theoretical pure modes with the off-diagonal elements set to zero.

To further verify the experimental and theoretical results and determine the chirality and polarization of these coupling modes, we performed additional micromagnetic simulations. Figure S5 shows that the simulations well reproduce the FMR experimental results at 7 K, as shown in Fig. 1(h) in main text. Based on the equilibrium positions of the magnetic moments at different magnetic fields, we can determine three special positions where the magnetic moments ($m_1$ and $m_2$) have a twofold rotational symmetry about *H*. According to the symmetry analysis, we can directly determine that there is no mode coupling between OM and AM at these three special angles, as shown in Figs. S5(b, d, and e). The spectra show that pure OM and AM cross completely without a coupling-induced frequency shift. However, when $H \parallel z$, the magnetic moments have an angle of 12 °with the twofold rotational symmetry axis, and coupling occurs between the OM and AM, as shown in Fig. S5(c). In contrast to the case of $H$ along the stronger in-plane anisotropy axis $b$, as shown in Fig. S5(d), the AM frequency is higher than the OM frequency at $\theta_a$ = 12°; thus, there is no frequency cross behavior, as shown in Fig. S5(e). However, when $H$ is within the $ab$-plane at an angle $\theta_{ab}$ = 45°, there exists mode coupling behavior between OM and AM with the characteristics of a coupling-induced frequency shift, as shown in Fig. S5(f).

To further connect the micromagnetic simulations with the quantum squeezing analysis, we compute the quadrature fluctuations directly from the simulated magnetization dynamics. Instead of applying a small coherent perturbation field, we change to use a finite temperature-induced fluctuating thermal field $\boldsymbol{B}_{\text{therm}}$ to excite spin-waves. In Mumax[3], classical thermal fluctuations in each cell are introduced through the stochastic field $\boldsymbol{B}_{\text{therm}}$ [13]

$$\boldsymbol{B}_{therm} = \boldsymbol{\eta}_{\boldsymbol{i}} \sqrt{\frac{2\alpha k_B T}{M_{sat}\, \gamma dV dt}}\ , \text{(S38)}$$

where $\boldsymbol{\eta}_i$ is a random vector from a standard normal distribution, $\alpha$ the damping parameter 0.01, $k_B$ the Boltzmann constant, $T$ the temperature, $\gamma$ the gyromagnetic ratio, $dV$ the cell volume and $dt$ the time step.

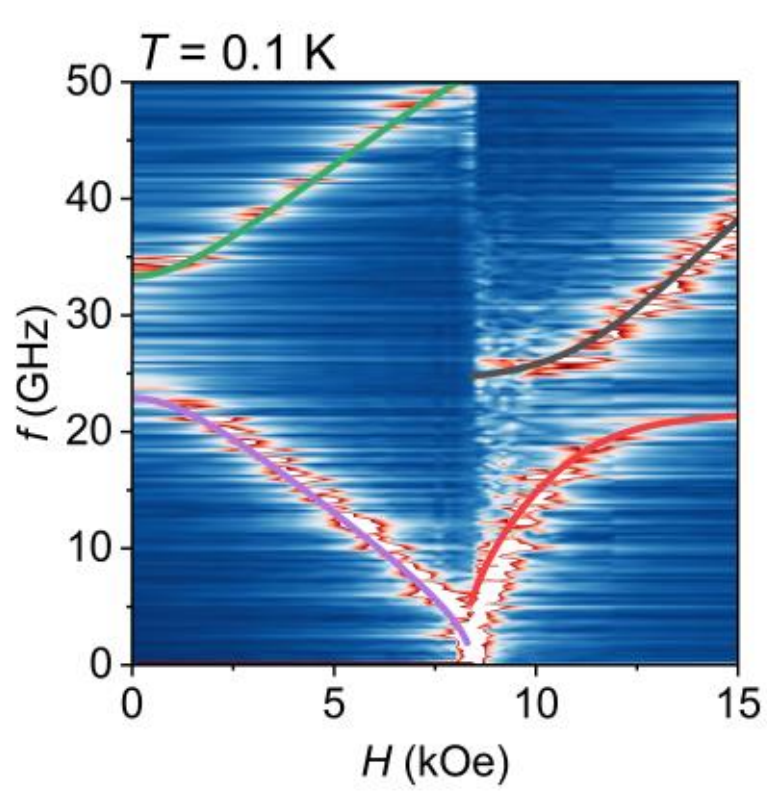

**Figure S6.** Theoretical dispersions of the dynamic modes and simulated results driven by a fluctuating thermal field ($T$ = 0.1 K).

From the time evolution of the spatially averaged magnetization $\boldsymbol{m}(t)$ over a duration of 4 ns, we obtained the *f-H* dispersion relations, as shown in Fig. S6, by performing a fast Fourier transform of $\boldsymbol{m}(t)$. The obtained frequency dispersion of the dynamical modes coincided with those obtained using a coherent perturbation field [Fig. S5], except for a relatively large background of noise. Using the coordinate systems defined in Fig. S4, we extracted the time evolution of the two quadratures $\boldsymbol{X}_1$ and $\boldsymbol{X}_2$ from these simulated magnetization trajectories. To quantify the thermal fluctuations, we evaluated the power spectral density of each quadrature and integrated it over frequency to obtain the corresponding variances. To eliminate background noise contributions, the integration was restricted to a narrow frequency window centered at the specific frequencies of the dynamical modes.

**Section V. Representative 26 K VNA-AFMR spectra and temperature-dependent fitting parameters**

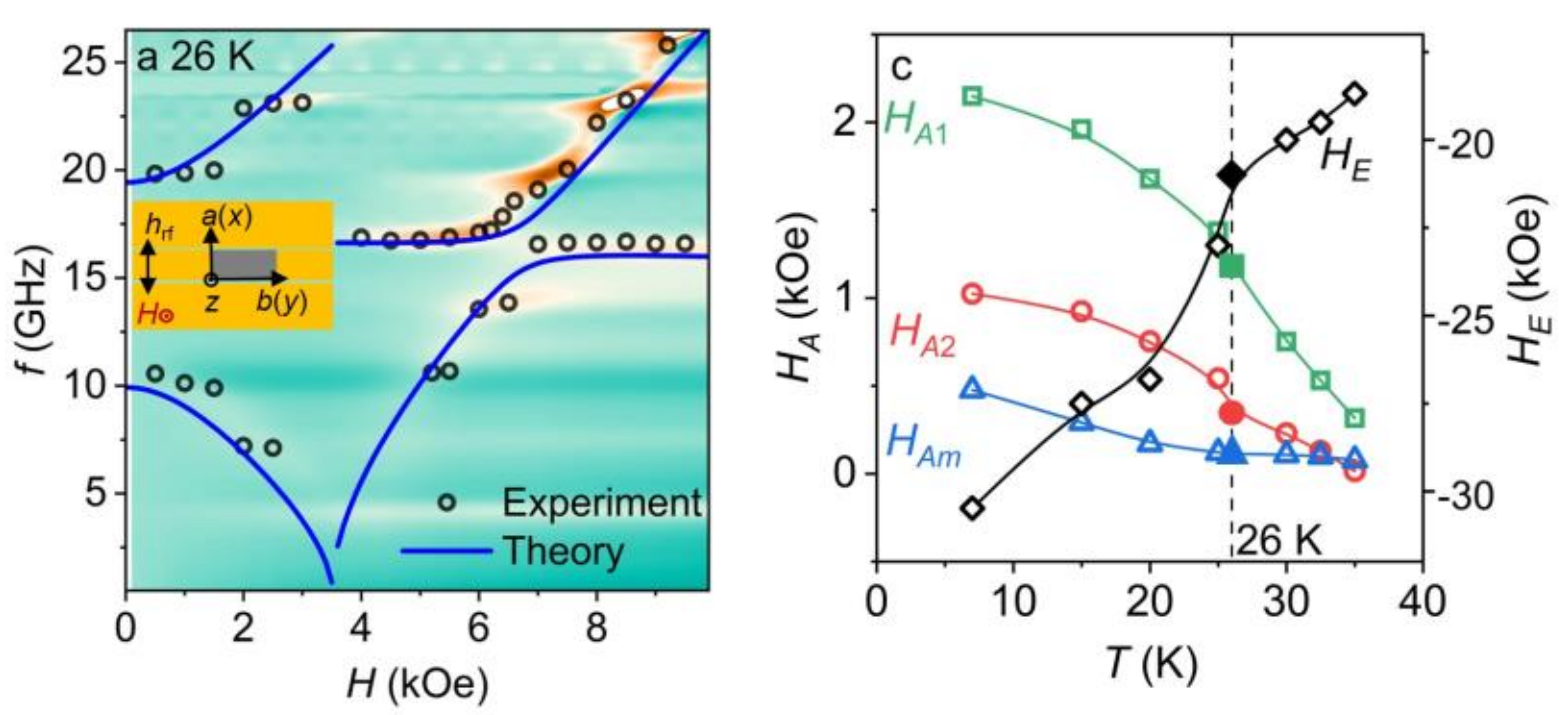

**Figure S7.** Representative VNA-AFMR $S_{21}$ spectra 26 K for $H \parallel z$ and summary of the temperature-dependent fitting parameters. (a) Color mapping of VNA-AFMR experimental results, extraction of the resonance frequency peaks, and theoretical dispersion at 26 K for $H \parallel z$. Insets: schematic of the experimental configuration. (b) Temperature dependence of the effective fields obtained by fitting differential FMR data in main text Fig. 1 (hollow symbols) and VNA-AFMR data (solid symbols) at 26 K in (a).

With the same experimental configuration as that for the differential FMR measurements shown in Fig. 1, we performed VNA-AFMR measurements at 26 K with the magnetic field applied along $H \parallel z$. The VNA-AFMR spectra are shown in Fig. S7(a). The extracted resonance frequencies were consistent with the theoretical dispersion curves, similar to the AFMR mapping in Fig. 4(g) of the main text. By fitting the dispersion using the theoretical model, we obtained the corresponding fitting parameters $H_{A1}$, $H_{A2}$, $H_{Am}$, and $H_E$ at 26 K. As summarized in Fig. S7(b), these fitting effective fields of 26 K are consistent with temperature-dependent trends independently determined from the differential FMR analysis in Fig. 1. Using these parameters obtained from the VNA-AFMR $S_{21}$ spectra at 26 K, we further calculated the angular-dependent dispersion curves using Eqs. (S9) and (S11). The calculated results are presented in solid lines in Fig. 4 of the main text.

Here, the pure AM and OM appear only when the two sublattice magnetic moments are symmetric with respect to the field, i.e., at $H \parallel b$ or $\theta_a = 96°$, as shown in Figs. 4(c) and (l). This absence of mode coupling originates from the twofold rotational symmetry of the magnetic moments $m_1$ and $m_2$ with respect to $H$, consistent with the zero value for the off-diagonal block matrix elements of the LLG dynamical matrix [Eq. (S9)]. At 26 K with the applied $H$ deviating from the spin direction ($AFM_c$ state) as shown in Figs. 4(d–g, h–k), the dynamic spectra cannot be decomposed into the RM-LM or AM-OM coupling cases above, but instead exhibit a more complex hybridization mode among RM, LM, AM, and OM. As a result, the mode-coupling strength cannot be analyzed using the same approach as before, and the VBSS and VRSS methods fail to diagnose the coupling-induced frequency shifts in these complex hybridization spectra. However, it remains feasible to approximately quantify the coupling strength $g$ at high-field dynamics using the AM-OM coupling method, while the RM-LM coupling method is more suitable for low-field dynamics. The frequency dispersions of the RM-LM or AM-OM coupling modes are illustrated by the solid and dashed lines in Fig. 4 of the main text.

**Section VI. Damping Analysis of Representative Differential AFMR Spectra**

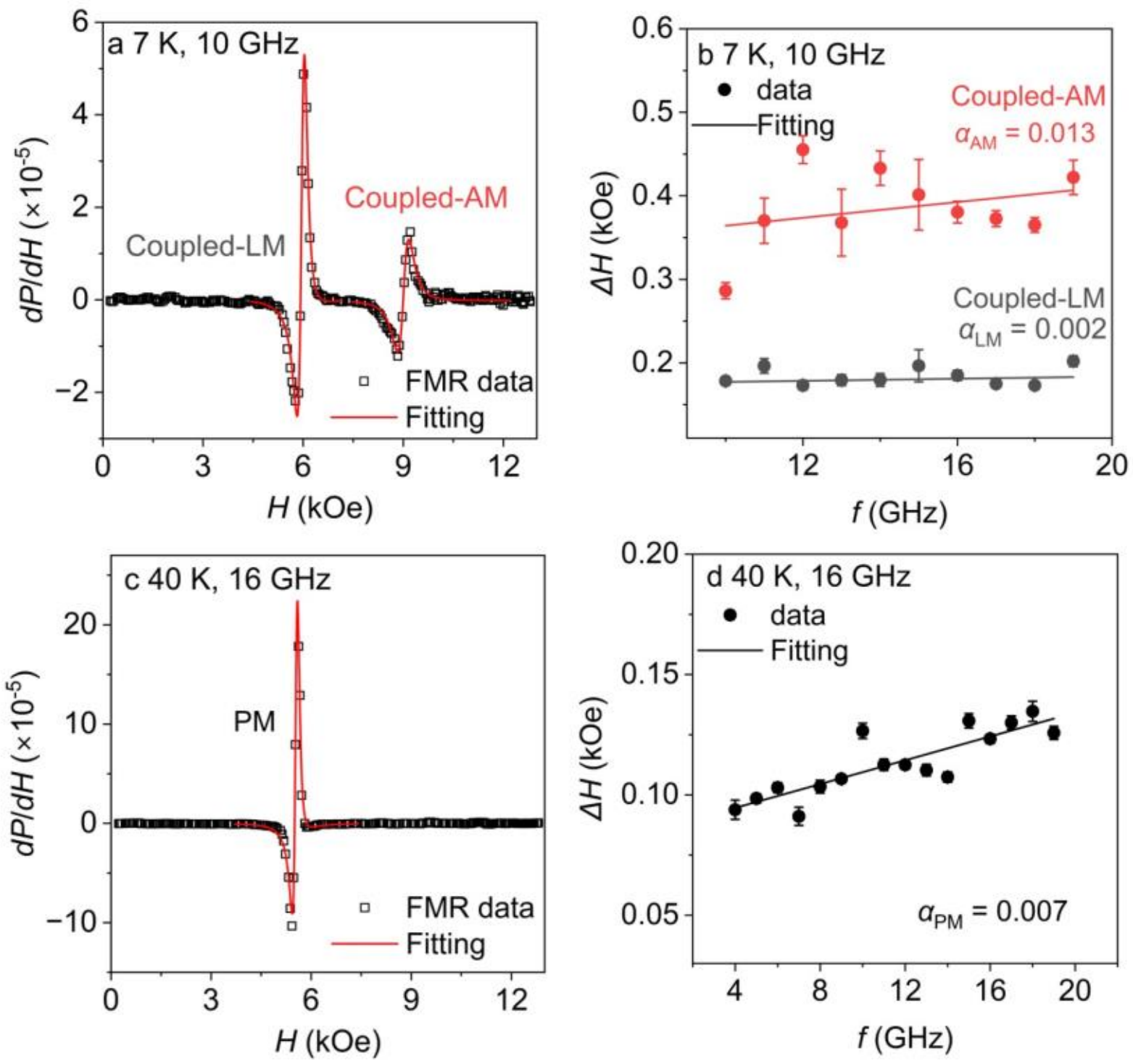

**Figure S8.** FMR spectra and damping analysis of $CrPS_4$ at 7 K and 40 K. (a) AFMR spectrum at 10 GHz and 7 K. The low-field peak corresponds to the coupled-LM, while the high-field peak corresponds to the coupled-AM. The solid lines are fits using differential Lorentzian functions, from which the half-linewidth $\Delta H$ is determined. (b) Frequency dependence of the half-linewidth $\Delta H$ for the coupled-LM and AM at 7 K. The solid lines are the fits used to determine the damping parameters. (c) AFMR spectrum at 16 GHz and 40 K. (d) Frequency dependence of $\Delta H$ for the paramagnetic mode (PM) at 40 K.

The cryogenic-temperature ferromagnetic resonance (FMR) spectra were acquired using a differential FMR measurement system that integrates the lock-in technique with a closed-cycle Gifford–McMahon refrigerator-based cryostat. A static magnetic field $H$ was modulated at an amplitude of 1~2 Oe and a frequency of 129.99 Hz via a pair of secondary Helmholtz coils. The resonance fields were determined by fitting the measured $dP/dH$ versus $H$ curves to a combination of symmetric and antisymmetric differential Lorentzian functions, as described in Ref. [14].

$CrPS_4$ is a low-damping van der Waals magnetic material. Taking the AFMR spectra shown in Fig. 1(a) at 40 K and Fig. 1(h) at 7 K as representative examples, we estimate the damping parameters of the coupled-AM mode, coupled-LM mode, and paramagnetic mode (PM). The Gilbert damping parameter $\alpha$ is conventionally extracted from the frequency dependence of the resonance half-linewidth $\Delta H$ using

$$\Delta H = \Delta H_0 + \frac{\alpha f}{\gamma}, \qquad (S39)$$

where $\Delta H_0$ accounts for inhomogeneous broadening and $\gamma$ is the gyromagnetic ratio. We applied this analysis to the AFMR spectra of $CrPS_4$ at 7 K and 40 K, as shown in Fig. S8. The half-linewidths were obtained by fitting the differential spectra $dP/dH$ with a linear combination of symmetric and antisymmetric Lorentzian derivative functions. From Eq. (S39), we obtain the following effective damping parameters: $\alpha_{\mathrm{LM}} = 0.002$ for the coupled-LM and $\alpha_{\mathrm{AM}} = 0.013$ for the coupled-AM. Additionally, the paramagnetic (PM) state with short-range magnetic order at $T = 40$ K, near the Néel temperature ($T_N$), yields $\alpha_{\mathrm{PM}} = 0.007$. These values show that $CrPS_4$, particularly its coupled-LM mode, exhibits low magnetic damping.